\documentclass[a4paper,11pt,nohyper]{JHEP3}
\def\d{\delta}

\def\t{\tau}

\def\S{\Sigma}

\def\del{\partial}              
   \let\d=\delta

 \let\t=\tau

\def\nn{\nonumber} \def\bd{\begin{document}} \def\ed{\end{document}}
\def\ds{\documentstyle} \let\fr=\frac \let\bl=\bigl \let\br=\bigr
\let\Br=\Bigr \let\Bl=\Bigl
\let\bm=\bibitem
\let\na=\nabla
\let\pa=\partial \let\ov=\overline
\newcommand{\be}{\begin{equation}}
\newcommand{\ee}{\end{equation}}
\def\ba{\begin{array}}
\def\ea{\end{array}}
\def\ft#1#2{{\textstyle{{\scriptstyle #1}\over {\scriptstyle #2}}}}
\def\fft#1#2{{#1 \over #2}}
\def\del{\partial}
\def\sst#1{{\scriptscriptstyle #1}}
 \def\oneone{\rlap 1\mkern4mu{\rm l}}
\def\ie{{\it i.e.\ }}
\def\via{{\it via}}
\def\semi{{\ltimes}}
\def\str{{\rm str}}
\def\Dm{{{D_{\sst{max}}}}}
\def\vac{ \left | 0 \right \rangle }
\def\kvac{ \left | k \right \rangle }

\def\sp{\; \; \;}

\def\bol{ \left | B (p^+) \right \rangle}
\def\bo1{ \left | B^0 (p^+) \right \rangle}

\def\bolt{ \left | B (p^+) \right \rangle_{\t}}

\def\boxl{ \left | B (x^-) \right \rangle}
\newcommand{\bea}{\begin{eqnarray}}
\newcommand{\eea}{\end{eqnarray}}

\def\<{ \langle }
\def\>{ \rangle }

\def\S{\Sigma}

\usepackage{amsmath,latexsym,amssymb,slashed}
\usepackage{graphicx}
\usepackage{psfrag}
\usepackage[latin1]{inputenc}
\usepackage{subfigure}

\renewcommand{\floatpagefraction}{0.6}
\renewcommand{\textfraction}{0.2}

\newcommand\ca{\mathcal{A}}
\newcommand\vp{\varphi}
\newcommand\beal{\begin{align}}
\newcommand\bbone{\ensuremath{\mathbbm{1}}}

\newcommand{\eq}[1]{\begin{equation}#1\end{equation}}
\newcommand{\spl}[1]{\begin{split}#1\end{split}}
\newcommand{\al}[1]{\begin{align}#1\end{align}}
\newcommand{\subeq}[1]{\begin{subequations}#1\end{subequations}}

\newcommand{\arXividhepth}[1]{\href{http://arxiv.org/abs/#1}arXiv:{\tt #1} [hep-th]}
\newcommand{\arXividother}[2]{\href{http://arxiv.org/abs/#1}arXiv:{\tt #1} [#2]}

\newcommand{\bg}[1]{\hat{#1}}
\newcommand{\wj}{\widetilde{J}}
\newcommand{\reo}{\mathrm{Re}~\!\omega}
\newcommand{\imo}{\mathrm{Im}~\!\omega}
\newcommand{\ads}{AdS_4}
\newcommand{\mcal}{\mathcal{M}}
\newcommand{\ccal}{\mathcal{C}}
\newcommand{\ncal}{\mathcal{N}}
\newcommand{\boxedeq}[1]{
\begin{equation}
\fbox{
\rule[0.7cm]{0pt}{0pt}
$#1$
\rule[-0.45cm]{0pt}{0pt}
}
\end{equation}
}

\def\d{\text{d}}

\def\slashchar#1{\setbox0=\hbox{$#1$}           
\dimen0=\wd0                                 
\setbox1=\hbox{/} \dimen1=\wd1               
\ifdim\dimen0>\dimen1                        
\rlap{\hbox to \dimen0{\hfil/\hfil}}      
#1                                        
\else                                        
\rlap{\hbox to \dimen1{\hfil$#1$\hfil}}   
/                                         
\fi}

\def\Re           {{\rm Re\hskip0.1em}}
\def\Im           {{\rm Im\hskip0.1em}}

\newcommand{\E}{\text{\tiny E}}

\newcommand{\tV}{{\widetilde{V}}}
\newcommand{\tH}{{\tilde{h}}}
\newcommand{\tm}{{{m}}}
\newcommand{\tmu}{{\tilde{\mu}}}
\newcommand{\trho}{{\tilde{\rho}}}
\newcommand{\tv}{{\tilde{v}}}
\newcommand{\calo}{\mbox{${\cal O}$}}
\newcommand{\cala}{\mbox{${\cal A}$}}
\newcommand{\dd}{\mathrm{d}}
\newcommand{\ra}{\rightarrow}
\newcommand{\calv}{\mbox{${\cal V}$}}
\newcommand{\calh}{\mbox{${\cal H}$}}
\newcommand{\calm}{\mbox{${\cal M}$}}
\newcommand{\abs}[1]{\left| #1 \right|}
\newcommand{\zetaa}{{\psi}}
\newcommand{\tr}{{\rm tr}\,}

\newcommand{\ky}[1]{{\color{blue}{#1}}}

\title{Entanglement entropy in top-down models}

\author{Peter A. R. Jones and Marika Taylor  \\

Mathematical Sciences and STAG Research Centre, University of Southampton, \\
Highfield, Southampton, SO17 1BJ, UK.

\bigskip
 E-mail:
 \email{p.jones@soton.ac.uk; m.m.taylor@soton.ac.uk}}

\abstract{We explore holographic entanglement entropy in ten-dimensional supergravity solutions. It has been proposed that entanglement entropy can be computed in such top-down models using minimal surfaces which asymptotically wrap the compact part of the geometry. We show explicitly in a wide range of examples that the holographic entanglement entropy thus computed agrees with the entanglement entropy computed using the Ryu-Takayanagi formula from the lower-dimensional Einstein metric obtained from reduction over the compact space. Our examples include not only consistent truncations but also cases in which no consistent truncation exists and Kaluza-Klein holography is used to identify the lower-dimensional Einstein metric. We then give a general proof, based on the Lewkowycz-Maldacena approach, of the top-down entanglement entropy formula. }

\begin{document}

\newcommand{\td}{\tilde}
 \newcommand{\bc}{\begin{center}}
 \newcommand{\ec}{\end{center}}
 \newcommand{\bfr}{\begin{flushright}}
 \newcommand{\efr}{\end{flushright}}
 \newcommand{\bfl}{\begin{flushleft}}
 \newcommand{\efl}{\end{flushleft}}
 \newcommand{\bt}{\begin{tabular}}
 \newcommand{\et}{\end{tabular}}

\section{Introduction}

In recent years there has been considerable interest in entanglement entropy and its holographic implementation, following the proposal of
\cite{Ryu:2006bv} that entanglement entropy can be computed from the area of a bulk minimal surface
homologous to a boundary entangling region. This proposal was proved for spherical entangling regions in conformal field theories in \cite{Casini:2011kv} and
arguments supporting the Ryu-Takayanagi prescription based on generalised entropy were given in \cite{Lewkowycz:2013nqa}.  Entanglement entropy has by now been computed in a wide range of holographic systems, see the review \cite{Takayanagi:2012kg}. General properties of the holographic entanglement entropy are reviewed in 
\cite{Headrick:2013zda}. 

The focus of this paper is on the computation of holographic entanglement entropy in top down systems. By ``top-down" we mean solutions of ten and eleven dimensional supergravity which are asymptotic to $AdS$ cross a compact space. In the context of phenomenological applications of holography, it is considered important to use top-down models wherever possible, to ensure that the quantities calculated are consistent.  Entanglement entropy is a novel computable for top down models and, following the pioneering works of  \cite{Klebanov:2007ws,Huijse:2011ef}, it can be used as an order parameter to characterise confinement and other phase transitions. 

The original Ryu-Takayanagi proposal \cite{Ryu:2006bv} is applicable to (asymptotically locally) anti-de Sitter spacetimes which are static. Given an entangling region on a spatial hypersurface of constant time in the boundary field theory, the entanglement entropy is computed holographically from the area $A$ of a bulk minimal surface of codimension two which is homologous to the boundary entangling region:
\be
S_{\rm RT} = \frac{A}{4 G_N} \label{rt}
\ee
where $G_N$ is the Newton constant. Note that the area of the minimal surface is computed in the Einstein frame metric. 
In the subsequent work \cite{Hubeny:2007xt} a covariant generalization of the Ryu-Takayanagi formula to non-static situations was proposed.

In this paper we will focus on entanglement entropy in top-down models, assuming that the solutions are globally static. (The latter is a reasonable assumption in many phenomenological models, in which holographic duals of Poincar\'{e} invariant field theories are being constructed, but the static assumption does exclude finite temperature and density models.)  

Consider a bulk solution which is asymptotic to $AdS_{d+1} \times X$ where $X$ is a compact space. Given an entangling region on a spatial hypersurface of the non-compact part of the boundary, then it has been suggested by \cite{Ryu:2006bv,Nishioka:2006gr} that the holographic entanglement entropy can be computed from the area of a codimension two minimal surface which asymptotically wraps the compact space $X$ and is homologous to the entangling region. Hence
\be
S_{\rm top-down} = \frac{{\cal A}}{4 \mathcal{G}_N} \label{hrt}
\ee
where ${\cal A}$ is the area of the minimal surface (in the Einstein frame metric) and $\mathcal{G}_{N}$ is the higher dimensional Newton constant. This prescription for the top-down entanglement entropy was used in \cite{Klebanov:2007ws} to explore phase transitions in top-down models. Other applications of the top-down prescription can be found in \cite{Faraggi:2007fu,Arean:2008az,Ramallo:2008ew,Bigazzi:2008qq,Bennett:2011va,Dey:2012hf,Kontoudi:2013rla,Bea:2013jxa,Aprile:2014iaa,Kim:2014yca,Bea:2015fja,Giusto:2015dfa}.

\bigskip

The purpose of this paper is to explore the relationship between \eqref{rt} and \eqref{hrt}. In particular, we will give strong evidence that the two formulae agree whenever we can uplift an asymptotically anti-de Sitter spacetime to a top-down solution. We will also give a proof that \eqref{hrt} indeed correctly calculates the holographic entanglement entropy in situations where consistent truncations of the top-down model do not exist, i.e. one does not know how to calculate the lower-dimensional Einstein metric.  
Our explicit examples focus primarily on asymptotically $AdS_5 \times S^5$ geometries, although the arguments and methodology could be straightforwardly generalized to other holographic dualities. 

As we review in section \ref{two}, the agreement between \eqref{rt} and \eqref{hrt} is manifest for top-down solutions which are globally direct products between an asymptotically locally $AdS$ geometry and a compact space $X$. The agreement between \eqref{rt} and \eqref{hrt} is far less obvious even in the context of consistent truncations of top down models to gauged supergravity. The map between the top-down Einstein metric and the lower-dimensional Einstein metric is quite complicated for consistent truncations, with warp factors depending non-trivially on both the lower-dimensional coordinates and on the position in the compact space, see for example \eqref{uplift}.  

In sections \ref{three} and \ref{four} we show that the top-down entanglement entropy computed via \eqref{hrt} indeed agrees with that computed using \eqref{rt} in consistent truncations to gauged supergravities and in consistent truncations involving massive vectors. The agreement involves non-trivial cancellations of warp factors depending on compact space coordinates. 

A generic asymptotically $AdS_5 \times S^5$ solution of ten-dimensional supergravity cannot be expressed as a solution of a five-dimensional theory which is a consistent truncation. For example, only special Coulomb branch solutions can be reduced to give gauged supergravity solutions (see examples in \cite{Freedman:1999gp,Freedman:1999gk}) and only a subgroup of LLM solutions \cite{Lin:2004nb} can be reduced to gauged supergravity solutions. However, in a finite region near the conformal boundary, one can always systematically reduce the ten-dimensional solutions over the sphere to obtain the five-dimensional Einstein metric as a Fefferman-Graham expansion; the reduction uses the methods of Kaluza-Klein holography developed in \cite{Skenderis:2006uy,Skenderis:2006di}.

In section \ref{kk-hol} we use Kaluza-Klein holography to compare the top-down entanglement entropy \eqref{hrt} with that obtained from the five-dimensional Einstein metric 
using \eqref{rt}, working up to quadratic order in the near boundary expansion. Even though the relationship between the five-dimensional and ten-dimensional Einstein metrics is extremely complicated (involving derivative field redefinitions), the expressions \eqref{rt} and \eqref{hrt} indeed agree. 

Entanglement entropy has also been computed for flavor brane solutions (used to describe flavors in the dual field theory), using both probe branes and backreacted (smeared) solutions. For probe branes, one can calculate the backreaction of the probe branes onto the lower-dimensional Einstein metric using Kaluza-Klein holography, see \cite{Jones:2015twa}, and show that this gives an equivalent answer to that obtained using \eqref{hrt}. Entanglement entropy for backreacted smeared solutions has previously been computed using \eqref{hrt}. In section \ref{six} we show that the same answer is obtained by extracting the lower-dimensional Einstein metric using Kaluza-Klein holography and applying \eqref{rt}, again confirming the matching between \eqref{rt} and \eqref{hrt}. 

\bigskip

Having established the agreement between \eqref{rt} and \eqref{hrt} in a number of examples, we give general arguments for why the formulae agree in section \ref{seven}, building on the approach of \cite{Lewkowycz:2013nqa}. In particular, assuming that the replica trick may be used, we can express entanglement in terms of partition functions for replica spaces. The latter can be computed holographically to leading order using the onshell action and therefore the equality of \eqref{rt} and \eqref{hrt} is essentially inherited from the equality of ten-dimensional and five-dimensional onshell actions. 

In section \ref{seven} we also give an alternative argument for the origin of \eqref{rt} and \eqref{hrt}, using the replica trick approach of 
 \cite{Lewkowycz:2013nqa} in combination with old results of Gibbons and Hawking on gravitational instanton symmetries \cite{Gibbons:1979xm}. The latter suggests that for generic entangling regions there may be additional contributions to the holographic entanglement entropy (even at leading order) if the circle direction used in the replica trick is non-trivially fibered over the boundary of the entangling region.  In practice one does not usually consider entangling regions such that the circle direction is non-trivially fibered but it would nonetheless be interesting to explore this situation further. 
 
\bigskip 
 
We conclude in section \ref{eight} by discussing the implications of our results for top-down holography and spacetime reconstruction. Extracting field theory data from a top-down solution is in general very subtle and computationally involved: one has to expand the ten-dimensional equations of motion perturbatively, and then use non-linear field redefinitions to obtain the effective five-dimensional equations of motion. Given the effective five-dimensional equations of motion and the asymptotic expansions of the five-dimensional fields, one can then read off field theory data using holographic renormalization \cite{Skenderis:2006uy,Skenderis:2006di}. We should note that these steps are required even to calculate quantities in the conformal vacuum: indeed non-linear field redefinitions between ten-dimensional and five-dimensional fields were first introduced in \cite{Lee:1998bxa} for the computation of three point functions in ${\cal N} = 4$ SYM. 

The lower-dimensional metric is a particularly important quantity for holography, as it relates to the dual energy momentum tensor. One needs to identify the lower-dimensional metric to compute one point functions and higher correlation functions of the stress energy tensor in the dual theory. The latter are in turn used in many contexts, including discussions of a theorems and also of energy correlations, following \cite{Hofman:2008ar}. Yet, as we review in section \ref{kk-hol}, the relation between the lower-dimensional metric and the ten-dimensional metric is very complicated. The matching of \eqref{rt} and \eqref{hrt} implies simple constraints relating the two metrics which can be used to check Kaluza-Klein holography calculations and perhaps even to deduce the lower-dimensional metric (see section \ref{six} for an example). 

There has been a great deal of interest in relating entanglement to the reconstruction of the holographic spacetime. Since \eqref{hrt} relates the entanglement entropy to minimal surfaces in the top-down geometry, entanglement implicitly knows about the compact part of the geometry. It would be interesting to explore further how entanglement can be used to understand the global structure of the ten-dimensional geometry.

\section{Entanglement Entropy for $AdS_5\times S^5$} \label{two}
We begin by reviewing the computation of entanglement entropy for a strip on the boundary of $AdS_5 \times S^5$ from both ten-dimensional and five-dimensional perspectives.  

Consider a strip $A$ defined by $x \in [0,l]$ on the boundary of $AdS_5$:
\begin{equation}
ds_5^2 =  \frac{1}{\rho^2} \left ( dx_{\mu} dx^{\mu} + {d\rho^2} \right ) 
\end{equation}
where $x^{\mu}=(t,x,y,z)$, the conformal boundary is at $\rho \rightarrow 0$, and we set the AdS radius to one throughout for convenience. 

To compute the entanglement entropy one calculates the area of a bulk codimension-2 minimal surface $\Sigma$ with boundary $\partial \Sigma =\partial A$:
\begin{equation}
S_5=\frac{1}{4G_5}\int_{\{ \Sigma | \partial \Sigma = \partial A  \}} d^3 \xi \sqrt{\textrm{det}\gamma_3}
\end{equation} 
where $\gamma_3$ is the induced metric on the minimal surface and $\xi_i$ $(i=1,2,3)$ are the worldvolume coordinates. Since the metric is static we work on a fixed-time slice $t=t_0$, and the surface is thus given by  $\Sigma=(t_0,x(\xi_i),y(\xi_i),z(\xi_i), \rho(\xi_i))$. By symmetry of the metric and boundary conditions it is clear that the surface cannot have non-trivial dependence on the $y,z$-directions, and (choosing static gauge to identify the $\xi_i$ with a subset of the spacetime coordinates) we can thus describe the minimal surface by an embedding of the form $x=x(\rho)$ or $\rho=\rho(x)$, where it is implicit that the surface extends in the $y,z$-directions. Taking $x=x(\rho)$ for concreteness the induced metric on $\Sigma$ is: 
\begin{equation}
ds^2_{ind}=  \frac{1}{\rho^2} \left[dy^2+dz^2+\left(x'^2+1\right)d \rho^2\right]
\end{equation}
and one can thus easily compute the entanglement entropy as:
\begin{equation}
S_5=\frac{V_2}{2G_5}\int_{\delta}^{\rho_0} \frac{d\rho}{ \rho^3 } \sqrt{x'^2+1 }
\end{equation}
where $V_2$ is the regularised area of $\partial A$, $\rho_0$ is the turning point of the surface, and $\delta$ is the UV cutoff. 

\bigskip

Now consider the calculation of the entanglement entropy from the ten-dimensional perspective, using the
ten-dimensional (Einstein) metric for $AdS_5 \times S^5$:
\begin{equation}
ds_{10}^2 = \frac{1}{ \rho^2 }\left(dx_{\nu} dx^{\nu} +d\rho^2\right) +d\theta^2+\textrm{cos}^2\theta d\Omega_3^2+\textrm{sin}^2\theta d\phi^2
\end{equation}
The proposed generalisation of the Ryu-Takayanagi prescription in this case is to calculate the area of a codimension two minimal surface $\Sigma$, now in the full ten-dimensional spacetime:
\begin{equation}
S_{10}=\frac{1}{4G_{10}}\int_{\{ \Sigma |\partial \Sigma = \partial A \}} d^8 \xi \sqrt{\textrm{det} \gamma_8}
\end{equation} 
where $\gamma_8$ is the induced metric on the minimal surface and $\xi_i$ $(i=1,...,8)$ are the worldvolume coordinates. 

Consider again the case of a strip on the boundary of the $AdS_5$ factor. In a similar fashion to before we can describe the corresponding minimal surface by an embedding of the form $x=x(\rho, \theta,\Omega_3,\phi)$, or $\rho=\rho(x, \theta,\Omega_3,\phi)$, or  $\theta=\theta(x, \rho,\Omega_3, \phi)$ etc., where we again have chosen static gauge, have assumed no dependence on the $y,z$-directions, and are working on a fixed-time slice $t=t_0$. However, due to the $S^5$ factor one must refine the boundary conditions to include the internal space. As before we take the boundary condition that the surface $\Sigma$ is anchored on $\partial A$, and consider further the condition that $\Sigma$ wraps the $S^5$ asymptotically. Alternative boundary conditions would describe different quantities in the dual field theory - see discussions on generalised entanglement entropy \cite{Mollabashi:2014qfa,Taylor:2015kda}. 

\begin{figure}
\begin{center}
\setlength{\unitlength}{0.50mm}
\includegraphics*[width=0.7\linewidth]{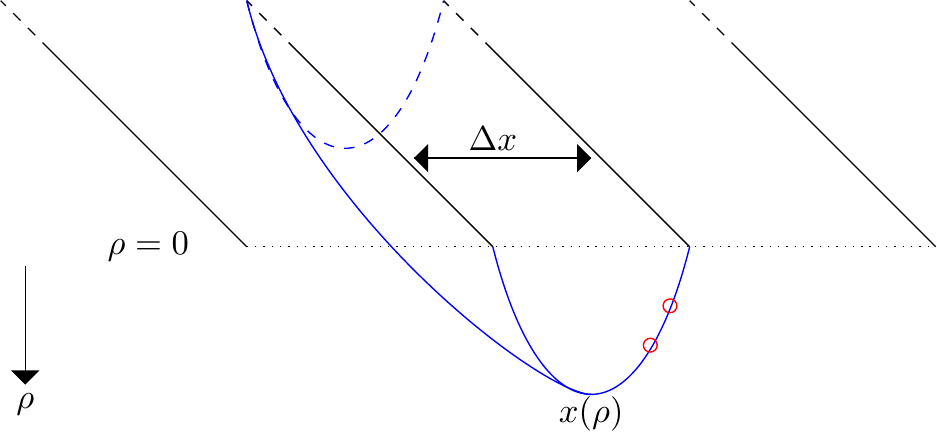}
\caption{The entangling surface for a slab boundary region - the conformal boundary is at $\rho \rightarrow 0$ and the minimal surface is described by $x(\rho)$. The minimal surface is a direct product of a codimension two surface in anti-de Sitter with the five sphere (the latter being indicated in red). }
\label{fig:slabsurface1}
\end{center}
\end{figure}

Since the $S^5$ is a maximally symmetric space and, importantly, the boundary conditions respect this symmetry, together with the fact that $AdS_5 \times S^5$ is a direct product, one can argue by symmetry as before that the minimal surface cannot depend non-trivially on the $S^5$ coordinates. Thus the embedding is of the form $x=x(\rho)$ or $\rho=\rho(x)$ as in the five-dimensional case, where it is now implicit that it both extends in the $y,z$-directions and wraps the $S^5$, see Figure~\ref{fig:slabsurface1}. Taking $x=x(\rho)$ for concreteness as before the induced metric on $\Sigma$ is just:
\begin{equation}
ds^2_{ind}=  \frac{1}{\rho^2} \left[dy^2+dz^2+\left(x'^2+1 \right)d \rho^2\right]+d\theta^2+\textrm{cos}^2\theta d\Omega_3^2+\textrm{sin}^2\theta d\phi^2
\end{equation}
The entanglement entropy is thus easily calculated to be:
\begin{equation}
S_{10}=\frac{V_2 V_{S^5}}{2G_{10}}\int_{\delta}^{\rho_0}  \frac{d\rho}{ \rho^3} \sqrt{x'^2+1 }
\end{equation}
which is identical to the 5-dimensional result since the Newton constants are related as $G_5=G_{10}/V_{S^5}$.

In the above example one hence obtains the same result for the entanglement entropy when computed from both the ten and five dimensional perspectives. This example had a particularly high level of symmetry, however, and it is not clear that the above equivalence should carry over to less trivial cases. 

The general problem one would like to study is the relationship between the entanglement entropy as calculated in a given downstairs metric and the entanglement entropy calculated in the uplifted solution, in cases where this uplift map is known or can be computed. Certain Coulomb branch geometries, which we study first in the following section, provide a good example of such a scenario, admitting a known ten-dimensional uplift which is not a simple product space, instead containing warp factors that depend on both the holographic radial coordinate and a sphere coordinate.

\section{Consistent Truncations of the Coulomb Branch} \label{three}

In this section we will consider particular Coulomb Branch solutions discussed in \cite{Freedman:1999gp,Freedman:1999gk}
that admit consistent truncations to solutions of five-dimensional gauged supergravity, and compare the entanglement entropy computed from five and ten dimensions. 

\subsection{Solutions with $SO(4)\times SO(2)$ Symmetry}

Let us discuss first Coulomb branch solutions which, from the ten-dimensional point of view, correspond to D3-branes being uniformly distributed on a disc of radius $\sigma$ in the transverse space. These supergravity solutions hence preserve $SO(4)\times SO(2)$ of the $SO(6)$ symmetry in the  $AdS_5 \times S^5$ solution. These Coulomb branch geometries admit consistent truncations to (a particular sector of) 5-dimensional gauged supergravity, with action given by:
\begin{equation}
I= \frac{1}{16 \pi G_5} \int d^5 x \left ( -\frac{1}{4}R+\frac{1}{2}(\partial \alpha)^2-\left ( \frac{g^2}{8}\left(\frac{\partial W}{\partial \alpha}\right)^2 -\frac{g^2}{3}W^2  \right ) \right )
\end{equation}
where $\alpha$ is a scalar field, $W$ is the superpotential and $g$ is the coupling constant. The five-dimensional Einstein frame metric for the solutions 
can be written as:
\begin{equation}
ds^2 = \lambda^2 w^2 \left ( dx_{\nu} dx^{\nu} +\frac{dw^2}{w^4 \lambda^6} \right ) \qquad
\lambda^6 = \left ( 1 + \frac{\sigma^2}{w^2} \right ), \label{5dmetric}
\end{equation}
which clearly reduces to an $AdS_5$ metric for $\sigma = 0$. (In the latter case the conformal boundary is at $w \rightarrow \infty$.)

Consider again a strip on the boundary defined by $x \in [0,l]$. As above we can describe the minimal surface by an embedding of the form $x=x(w)$ or $w=w(x)$. Taking $x=x(w)$ the induced metric on the surface is easily calculated to be:
\begin{equation}
ds^2_{ind}= \lambda^2 w^2 \left[dy^2+dz^2+\left(x'^2+\frac{1}{w^4 \lambda^6}\right)d w^2\right]
\end{equation}
where $x'\equiv dx/dw$, and thus one finds:
\begin{equation}
\sqrt{\textrm{det}\gamma}=\lambda^3 \rho^3 \sqrt{x'^2+\frac{1}{w^4 \lambda^6}}
\end{equation}
The entanglement entropy for the slab is thus:
\begin{equation}
S=\frac{V_2}{2 G_5}\int_{w_0}^\Lambda dw \lambda^3 w^3 \sqrt{x'^2+\frac{1}{w^4 \lambda^6}} \label{EE5dx}
\end{equation}
where $w_0$ is the turning point of the minimal surface and $\Lambda$ is the UV cutoff. 

\bigskip

The five-dimensional metric in (\ref{5dmetric}) can be uplifted to the following ten-dimensional Einstein frame metric  \cite{Freedman:1999gp,Freedman:1999gk}:
\begin{equation}
ds_{10}^2 = \Delta^{-2/3} ds^2 + ds_K^2, \label{uplift}
\end{equation}
where the warp factor $\Delta$ depends both on the holographic radial coordinate and on one of the sphere coordinates, while $ds_K^2$ is a metric on a warped sphere. Explicit expressions for these quantities are:
\begin{equation}
\Delta^{-2/3} = \frac{\zeta}{\lambda^2} \hspace{15mm} \zeta = (1 + \frac{\sigma^2}{w^2} \cos^2 \theta)
\end{equation}
\begin{equation}
ds_K^2 =\frac{1}{\zeta} \left ( \zeta^2 d \theta^2 + \cos^2 \theta d\Omega_3^2 + \lambda^6 \sin^2 \theta d \phi^2 \right ). 
\end{equation}
Note that $\zeta,\lambda \rightarrow 1$ as $w \rightarrow \infty$ and thus the solution is indeed asymptotically $AdS_5 \times S^5$. To compute the entanglement entropy for the strip  we now proceed as before, with the additional boundary condition that the minimal surface wraps the $S^5$ asymptotically. 

However, in the present case there are non-trivial warp factors that mix the holographic radial coordinate $w$ and the sphere coordinate $\theta$. Thus, although we can continue to assume the minimal surface has trivial dependence on $\Omega_3$ and $\phi$, we can no longer a priori assume that the minimal surface has trivial dependence on $\theta$. We thus may assume an embedding of the form $x=x(w,\theta)$, or $w=w(x,\theta)$ or $\theta=\theta(x,w)$. Choosing $x=x(w,\theta)$ (as the boundary conditions will be clearest in this choice) one calculates the induced metric to be:
\bea
ds^2_{ind} &=&  \zeta w^2 \left[dy^2+dz^2+\left(x'^2+\frac{1}{w^4 \lambda^6}\right)d \rho^2\right]+ 2 \zeta w^2 x'\dot{x} d\rho d\theta+\zeta(1+\dot{x}^2 w^2)d\theta^2 \\
&&  \qquad \qquad  +\frac{\textrm{cos}^2\theta}{\zeta}d\Omega_3+\frac{\lambda^6}{\zeta}\textrm{sin}^6\theta d\phi \nn 
\eea
where $\dot{x}\equiv dx/d\theta$. One thus finds that the ten-dimensional entanglement functional is 
\be
S = \frac{1}{4 G_{10}} \int d^8 x \sqrt{\textrm{det} \gamma}
\ee
where 
\begin{equation}
\sqrt{\textrm{det}\gamma}=\sqrt{\textrm{det}g_{\Omega_3}}\textrm{cos}^3\theta \textrm{sin} \theta \lambda^3 w^2  \left[\left(\rho^2 x'^2+\frac{1}{w^2 \lambda^6} \right)(1+\dot{x}^2 w^2)-\dot{x}^2x'^2 w^4\right]^{1/2} \label{lag}
\end{equation}
Notice that all factors of $\zeta$, which depend on the sphere coordinate $\theta$, cancel out and the spherical prefactors combine to become $\sqrt{\textrm{det}g_{\Omega_5}}$. The only additional dependence on the spherical coordinates thus comes through the fact that $x(w,\theta)$ depends on $\theta$: 
\begin{equation}
\sqrt{\textrm{det}\gamma}=\sqrt{\textrm{det}g_{\Omega_3}}\textrm{cos}^3\theta \textrm{sin} \theta \lambda^3 w^2  \sqrt{w^2 x'^2+\frac{\dot{x}^2 }{ \lambda^6}+\frac{1}{w^2 \lambda^6}} \label{EL2var}
\end{equation}
One can immediately make an interesting observation. The equations of motion admit the solution $\dot{x} = 0$, since the action is quadratic in $\dot{x}$ and $\theta$ does not appear explicitly in the non-trivial square root part of the action functional.  For solutions in which $\dot{x} = 0$, the entanglement entropy is thus:
\begin{equation}
S=\frac{V_2 V_{S^5}}{4G_{10}}\int_{w_0}^\Lambda dw \lambda^3 w^3 \sqrt{x'^2+\frac{1}{w^4 \lambda^6}} 
\end{equation}
i.e. identical to (\ref{EE5dx}) since $G_5=G_{10}/V_{S^5}$. 

Although one can thus consistently set $\dot{x}=0$ to obtain a solution to the ten-dimensional equation of motion, it remains to show that this is indeed the minimal solution. This can be done using the radial Hamiltonian formalism as follows. One first assumes a given $\theta$-independent solution of the equations of motion and then considers $\theta$-dependent perturbations to this background. By computing the Hamiltonian one then shows that these perturbations lead to a larger Hamiltonian and thus the minimal solution (at least perturbatively) is indeed the one that is independent of $\theta$. The independence of the minimal surface on the compact coordinates is a point we will return to in section \ref{section52}.

\subsection{Other Coulomb Branch Solutions}
Similar conclusions can be reached for other consistent truncations of Coulomb branch solutions with different symmetries. In \cite{Freedman:1999gp,Freedman:1999gk} they also consider solutions with $SO(3)\times SO(3)$ and $SO(5)$ symmetry in addition to the $SO(4)\times SO(2)$ solution considered above, corresponding to various symmetric distributions of D3-branes. The $SO(3)\times SO(3)$ case has the following 10d metric:
\begin{equation}
ds_{10}^2=\zeta w^2 \lambda \left( dx_{\mu}^2 + \frac{dw^2}{w^4 \lambda^6} \right) +\frac{1}{\lambda \zeta} \left(\zeta^2 d\theta^2 + \textrm{cos}^2 \theta d\Omega_2^2+\lambda^4 \textrm{sin}^2 \theta d\tilde{\Omega}_2^2 \right)
\end{equation}
\begin{equation}
\Delta^{-2/3}=\frac{\zeta}{\lambda}
\end{equation}
while the $SO(5)$ case has the following 10d metric:
\begin{equation}
ds_{10}^2=\frac{\zeta w^2 }{\lambda^3} \left( dx_{\mu}^2 + \frac{dw^2}{w^4 \lambda^6} \right) +\frac{\lambda^3}{ \zeta} \left(\zeta^2 d\theta^2 + \textrm{cos}^2 \theta d\Omega_4^2 \right)
\end{equation}
\begin{equation}
\Delta^{-2/3}=\frac{\zeta}{\lambda}
\end{equation}
In both cases the definitions of $\lambda$ and $\zeta$ are as before, and the expression for $\Delta$ shows the relationship between the ten-dimensional and five-dimensional metrics c.f. (\ref{uplift}). Given what has been deduced from the $SO(4) \times SO(2)$ case previously, it is immediate that the same equivalence will occur in these cases, since the factors of $\zeta$ cancel in the determinant and indeed one can explicitly check that the powers of $\lambda$ come out the same in the two cases.

\section{Consistent Truncations with Massive Vector Fields} \label{four}

In this section we consider the entanglement entropy for particular backgrounds which admit consistent truncations with massive vector fields, as discussed in \cite{Maldacena:2008wh}. Consider again type IIB supergravity but now with the metric, the dilaton $\Phi$, the 5-form $F_5$, and the 3-form $H=\textrm{d}B$ switched on. 
Our conventions for the action in Einstein frame are
\begin{equation}
I=\frac{1}{16 \pi G_{10}}\int d^{10}x \sqrt{-g_{10}}\left[R-\frac{1}{2}\partial_A \Phi \partial^A \Phi -\frac{1}{2 \cdot 3!}e^{-\Phi}H_{ABC}H^{ABC}-\frac{1}{2 \cdot 5!} F_{(5)}^2 \right] \label{iib}
\end{equation}
where as usual we need to impose in addition the self-duality constraint on $F_5$. 

Now consider the following ansatz for the ten-dimensional fields:
\begin{gather}
ds_{10}^2=e^{-\frac{2}{3}(4U+V)}ds^2_M+e^{2U}ds^2_{B_{KE}}+e^{2V}\eta^{2} \label{uplift1} \\
B=A \wedge \eta +\theta \omega \label{uplift2}\\
F_5=4e^{-4U-V}(1+ \star)vol_M \label{uplift3}
\end{gather}
where $M$ is the 5-dimensional spacetime with metric $ds^2_M$ and volume form $vol_M$.
Furthermore, $ds^2_{B_{KE}}+\eta^2$ is a Sasaki-Einstein metric c.f. the representation of $S^5$ as a $U(1)$ fibration over $\mathbb{CP}^2$. The scalars $U$, $V$ and $\Phi$ are taken to be functions on $M$, as is the one-form $A$. Expressions for the quantities $\theta$ and $\omega$ will not be important in what follows but may be found in \cite{Maldacena:2008wh}. 

Reducing the field equations over the internal space, one obtains equations of motion which may be derived from the following 5-dimensional action for the fields $(g_5, U, V, \Phi, A)$: 
\begin{equation}
\begin{split}
I=\frac{1}{16 \pi G_5}\int d^{5}x \sqrt{-g_{5}} \Big[R +24e^{-u-4v}-4e^{-6u-4v}-8e^{-10v}-5 (\partial u)^2 -\frac{15}{2} (\partial v)^2  \\ -\frac{1}{2} (\partial \Phi)^2 -\frac{1}{4}e^{-\Phi+4u+v}F_{mn}F^{mn}-4e^{-\Phi-2u-3v}A_m A^m \label{redaction} \Big]
\end{split}
\end{equation}
where $F=dA$,  $u=\frac{2}{5}(U-V)$ and $v=\frac{4}{15}(4U+V)$. It was shown in \cite{Maldacena:2008wh} that this reduction is consistent i.e. any solution of the resulting five-dimensional equations of motion can be uplifted to a solution of type IIB supergravity using the map (\ref{uplift1})-(\ref{uplift3}). 

Note that from the reduced action (\ref{redaction}) one finds that the mass of the vector field $A$ around the $AdS_5$ background is $m^2=8$, showing these solutions are indeed associated with massive vector fields. As is clear from (\ref{uplift1})-(\ref{uplift2}) however, this vector field does not appear in the ten-dimensional metric but instead appears in the ten-dimensional two-form field and thus it does not directly contribute to the ten-dimensional entanglement entropy. 

We can immediately compute the ten-dimensional entanglement entropy, which as before is given by:
\begin{equation}
S_{10}=\frac{1}{4G_{10}}\int_{\{ \Sigma |\partial \Sigma = \partial A \}} d^8 \xi\sqrt{\textrm{det} \gamma_8}
\end{equation} 
where implicitly we work with the metric in Einstein frame. One can now immediately obtain the ten-dimensional entanglement entropy for an arbitrary entangling region, only assuming that we again work on a fixed time slice and that the entangling surface wraps the internal space asymptotically. Since the warp factors in the metric do not depend at all on the internal directions the entangling surface will therefore also wrap the internal space deep in the bulk. Since the entangling surface is consequently codimension two with respect to the five-dimensional spacetime $M$ one trivially obtains:
\begin{equation}
\sqrt{\gamma_8}=(e^{-\frac{2}{3}(4U+V)})^{\frac{3}{2}} (e^{2U})^{\frac{4}{2}}(e^{2V})^{\frac{1}{2}}\sqrt{\gamma_5}\hspace{1mm}vol_{SE}=\sqrt{\gamma_5}\hspace{1mm}vol_{SE}
\end{equation}
where $vol_{SE}$ is the volume form on the internal space, and thus it is immediate that the entanglement entropy as computed from ten dimensions will be equivalent to the five-dimensional entanglement entropy. 

A particular example of interest in this solution class is given by backgrounds with non-relativistic scaling symmetries, in particular the Schr\"{o}dinger backgrounds discussed in \cite{Maldacena:2008wh}. These are deformations of $AdS$ that have a metric that can be written in the following form:
\begin{equation}
ds^2_{M_z}=-b^2 r^{2z} (dx^+)^2+\frac{dr^2}{r^2}+r^2\left(-dx^-dx^++dx^2+dy^2\right) \label{schrodinger}
\end{equation}
where $x^{\pm}$ are lightcone coordinates, $z$ is the dynamical exponent and $b$ is a parameter that characterizes the deformation from $AdS_5$. This metric is a solution to the equations of motion one obtains from the following action:
\begin{equation}
S=\frac{1}{16 \pi G_{D+3}} \int d^{D+2}xdr \sqrt{-g}\left(R-2\Lambda-\frac{1}{4}F_{mn}F^{mn}-\frac{m^2}{2}A_{m}A^{m}\right) \label{schaction}
\end{equation}
where the vector field solution is $A_+ \propto r^z$, provided that $\Lambda=-(D+1)(D+2)/2$ and $m^2=z(z+D)$. 

One can check that the metric (\ref{schrodinger}) for $z=2$ (and $D=2$) together with $U=V=\Phi=0$ and $A_+=b  r^2$ is a solution to the equations of motion one derives from (\ref{redaction}) - indeed, (\ref{redaction}) reduces to  (\ref{schaction}) under these conditions, where in the present case $m^2=8$ as expected. Checking explicitly the equivalence of the ten-dimensional and five-dimensional entanglement entropies is trivial in this case since all the warp factors in (\ref{uplift1}) evaluate to one and thus the metric is a simple product space. Note that an identical analysis can be performed for consistent truncations that have vector fields with mass $m^2=24$ found in \cite{Maldacena:2008wh}, and the equivalence between the ten-dimensional and five-dimensional entanglement entropy carries over in the same way in such cases. 

\section{Kaluza-Klein Holography} \label{kk-hol}

A generic ten dimensional supergravity solution which is asymptotic to $AdS_5 \times S^5$ cannot be expressed as the uplift of a five dimensional supergravity solution. However, in the vicinity of the conformal boundary the ten-dimensional solution can always be expressed as a perturbation of $AdS_5 \times S^5$. Dual field theory data can be expressed in terms of these perturbations using the method of Kaluza-Klein holography \cite{Skenderis:2006uy,Skenderis:2006di}, as we now review. 

Let us express the $AdS_5 \times S^5$ 
metric as 
\begin{equation}
ds^2 =  g^{o}_{AB} dx^A dx^B \equiv \frac{1}{\rho^2} \left ( d \rho^2 + dx^{\mu} dx_{\mu} \right )  + d \Omega_5^2
\end{equation}
with the five form flux being 
\begin{equation}
F = F^o \equiv \eta_{AdS_5} + \eta_{S^5} 
\end{equation}
where $\eta$ denotes the volume form. The Einstein metric of a solution of the type IIB equations which is a deformation of $AdS_5 \times S^5$ can therefore be expressed as 
\begin{equation}
g_{AB} = g^o_{AB} + h_{AB}.
\end{equation} 
The metric fluctuation can always be decomposed in terms of spherical harmonics on the sphere. The metric fluctuations are decomposed as
\bea
h_{mn} &=& \sum h_{mn}^I Y^I; \\
h_{ma} &=& \sum \left ( B^{I_v}_{m} Y^{I_v}_{a} + b_m^{I} D_{a} Y^I \right ); \nn \\
h_{(ab)} &=& \sum \left ( \phi^{I_t} Y^{I_t}_{(ab)} + \psi^{I_v} D_{(a} Y^{I_v}_{b )} + \chi^I D_{(a} D_{b)} Y^I \right ); \nn \\
h^{a}_{a} &=& \sum \pi^I Y^I \nn
\eea
where $Y^I$ are scalar harmonics, $Y^{I_v}_a$ are vector harmonics and $Y^{I_t}_{(ab)}$ are symmetric traceless tensor harmonics; $D_a$ denotes the covariant derivative. We will not need explicit forms for the spherical harmonics in what follows but note that the defining equations are:
\bea
\Box Y^I &=& \Lambda^I Y^I \qquad \Lambda^I = - k(k+4) \qquad k=0,1,2,\cdots \\
\Box Y^{I_v}_a &=& \Lambda^{I_5} Y^{I_v}_a \qquad \Lambda^{I_v} = - (k^2 + 4k -1) \qquad k =1,2,\cdots \nn \\
\Box Y^{I_t}_{(ab)} &=& \Lambda^{I_{t}} Y^{I_t}_{(ab)} \qquad \Lambda^{I_t} = - (k^2 + 4k -2 ) \qquad k=2,3,\cdots \nn
\eea
where $\Box$ is the D'Alambertian and $D^a Y_{a}^{I_v} = D^{a} Y_{(ab)}^{I_t} = 0$. The spherical harmonic labels denote both the degree of the harmonic and additional quantum numbers, i.e. charges under the Cartan of $SO(6)$. 

The fluctuations are not all independent, as some of the modes are diffeomorphic to each other or to the background. To derive the spectrum around $AdS$ it is usual to impose a gauge fixing condition such as the de Donder-Lorentz gauge
\be
D^a h_{(ab)} = D^a h_{am } = 0
\ee
which sets to zero $b_{m}^{I}$, $\psi^{I_v}$ and $\chi^I$. The remaining modes $h^{I}_{mn}$, $B^{I_v}_{m}$, $\phi^{I_t}$ and $\pi^I$ are then related to tensor, vector and scalar fields in five dimensions. Although this gauge choice is very convenient for deriving the spectrum, it can be less useful when analysing a generic solution, as typically such solutions will not naturally be expressed in this gauge. Instead of gauge fixing the symmetry, one can instead derive gauge invariant combinations of the fluctuations; the latter are the five-dimensional fields \cite{Skenderis:2006uy,Skenderis:2006di}. 

Working to linear order in the perturbations the five-dimensional Einstein metric $g^{5}_{mn} = g^o_{mn} + H_{mn}$ is related to the ten-dimensional metric perturbations given above as
\be
H_{mn} = h_{mn}^{0} + \frac{1}{3} \pi^{0} g^{o}_{mn}, \label{linear-ein}
\ee 
i.e. it depends only on the zero mode of the tensor perturbation and the breathing mode on the sphere. The origin of the second term is the Weyl rescaling needed to bring the five dimensional metric into Einstein frame.

Working to quadratic order in the perturbations, the expression for the five-dimensional metric in terms of the ten-dimensional metric perturbations is considerably more complicated and indeed it has not been worked out in generality. At quadratic order the schematic form of the appropriately gauge-invariant metric perturbation is
\be
h_{mn} = h_{mn}^0 + \frac{1}{3} \pi^0 g^{o}_{mn} + h_{(2) mn} 
\ee
where $h_{(2) mn}$ is quadratic in perturbations. 

For example, for modes associated with the scalar spherical harmonics the quadratic contributions are \cite{Skenderis:2006uy}
\bea
h_{(2) mn} &=& - \sum_I z(k) \left ( \frac{1}{2} \Lambda^I (\chi^I \hat{h}^I_{mn} + \frac{1}{2} D_{m} \chi^I D_n \chi^I)  \right . \\
&& + \left . D_{m} \hat{b}^{pI} \hat{h}^I_{n p} +  D_{n} \hat{b}^{pI} \hat{h}^I_{m p} + \hat{b}^{pI} D_{p} \hat{h}^I_{mn} + D_{m} \hat{b}^{pI} D_{n} \hat{b}^{I}_{p} + \hat{b}^{pI} \hat{b}^{I}_{p} g^{o}_{mn} - \hat{b}^{I}_{m} \hat{b}^{I}_{n} \right ) \nn
\eea
where 
\bea
\hat{b}^{I}_{m} &=& b^{I}_{m} - \frac{1}{2} D_{m} \chi^I \\
\hat{h}^{I}_{mn} &=& h^{I}_{mn} - D_{m} b^{I}_{n} - D_{n} b^{I}_{m} \nn 
\eea
are gauge invariant combinations at linear order in the fluctuations. Note that if we work in de Donder-Lorentz gauge $h_{(2) mn}  = 0$. 

It is important however to note that $h_{mn}$, while appropriately gauge invariant with respect to the ten-dimensional symmetries and transforming as a five-dimensional metric, is still not the five-dimensional Einstein metric fluctuation. The combination $h_{mn}$ satisfies an Einstein equation 
\be
( {\cal L}_{E} + 4 ) h_{mn} = T_{(2) mn}
\ee
where ${\cal L}_E$ is the usual linearized Einstein operator and  the effective stress energy tensor is $T_{(2) mn}$. This effective stress energy tensor is quadratic in the fluctuations but  involves derivative interactions. For example, terms quadratic in the fields $\pi^I$ have the general structure
\be
T_{(2) mn} = \sum_I  \left ( a_I D_{m} D_{p} D_{r}  \pi^I D_{n} D^{p} D^r \pi^I   + b_{I} D_{m} D_{p} \pi^I D_{n} D^{p} \pi^I + \cdots \right )
\ee
with certain coefficients $(a_I,b_I,\cdots)$. The effective five-dimensional action does not contain derivative interactions, and therefore the five-dimensional fields must be related to ten-dimensional fields by non-linear field redefinitions, as first noted in \cite{Lee:1998bxa}. In particular the five-dimensional Einstein metric perturbation $H_{mn}$ is related to the metric fluctuation $h_{mn}$ as 
\be
H_{mm} = h_{mn} + \sum_{I} \left ( A_{I} D_{m} D_{p} \pi^I D_{n} D^{p} \pi^I + B_{I} D_{m} \pi^I D_{n} \pi^I + \cdots \right )
\ee
where again the coefficients $(A_I,B_I,\cdots)$ are computable. Thus the explicit form of the five-dimensional Einstein metric is extremely complicated at quadratic order since it involves infinite sums with coefficients $(A_I,B_I,\cdots)$ which are very arduous to compute; see  \cite{Skenderis:2006uy} for explicit expressions. 

\subsection{General Coulomb Branch solutions} \label{gcb}

As an example of solutions which can be understood using Kaluza-Klein holography, we consider general Coulomb branch solutions i.e. solutions that do not necessarily admit a consistent truncation. The metric for such solutions takes the following form:
\begin{equation}
ds^2=H(y)^{-1/2}dx_{\mu} dx^{\mu}+H(y)^{1/2}dy_{i} dy^{i} \label{CBmetric}
\end{equation}
where $x_{\mu}$ are the brane directions and $y_i$ are transverse directions, and $H(y)$ is a harmonic function on $R^6$. Near the conformal boundary the harmonic function takes the form
\be
H = \frac{L^4}{r^4} \left ( 1 + \sum_{k \ge 2} \frac{a_{k}^I Y_k^I}{r^k} \right )
\ee
where we have written
\be
dy_{i} dy^{i} = dr^2 + r^2 d \Omega_5^2
\ee
while $Y_k^I$ are scalar harmonics of degree $k$ on $S^5$ and $a_k^I$ are coefficients defining the brane distribution. Implicitly we have taken the decoupling limit of the brane solution, i.e. dropped the constant term in the harmonic function. 

We can now express the Coulomb branch metric asymptotically as a perturbation of $AdS_5 \times S^5$. The background asymptotes to  
\be
ds^2 = g^o_{AB} dx^A dx^B = \frac{r^2}{L^2} dx_{\mu} dx^{\mu} + \frac{L^2}{r^2} dr^2 + L^2 d \Omega_5^2
\ee
To match with earlier conventions we set $L^2=1$ (the curvature radius can be reinstated in final formulae if required). Near the conformal boundary
\be
g_{AB} = g^o_{AB} + h_{AB}
\ee
where working to linear order in the coefficients $a_n^I$ we can read off:
\bea
h_{\mu \nu} &=& - \sum_{k \ge 2} \frac{a_k^I Y_k^I}{2 r^{k-2}} \eta_{\mu \nu}; \\
h_{rr} &=&  \sum_{k \ge 2} \frac{a_k^I Y_k^I}{2 r^{k+2}}; \nn \\
h_{ab} &=&   \sum_{k \ge 2} \frac{a_k^I Y_k^I}{2 r^{k}} g^o_{ab}. \nn
\eea
Hence the non-zero perturbations are
\be
\pi^I = \frac{5 a_k^I}{2 r^k} \label{pi-ki}
\ee
and
\be
h_{\mu \nu}^I = - \frac{a_k^I}{2 r^{k-2}} \qquad
h_{rr}^I = \frac{a_k^I}{2 r^{k+2}},
\ee
for harmonics of degree $k \ge 2$. 

These perturbations are consistent with the diagonalised equations of motion at linear order found in \cite{Kim:1985ez}.
Let
\be
\pi^I = 10 k \epsilon s^I 
\ee 
where $\epsilon$ is a small parameter and $k$ is the degree of the associated spherical harmonic with $k \ge 2$. The equation of motion for $s^I$ is
\be
\Box s^I = k (k-4) s^I \label{eom-s}
\ee
where $\Box$ is the d'Alambertian in $AdS_5$. 

The supergravity field equations at linear order then imply that such perturbations are necessarily accompanied by 
\bea
h^{I}_{mn} = \epsilon h^I_{(1) mn} &=& \epsilon \left ( \frac{4}{(k+1)} D_{(m} D_{n)} s^I - \frac{6k}{5} s^I g^{o}_{mn} \right) \\
&=&  \epsilon \left ( \frac{4}{(k+1)} D_{m} D_{n} s^I - \frac{2k}{k+1} (k-1) s^I g^{o}_{mn} \right) \nn
\eea
If one switches on only these modes at linear order, as in the Coulomb branch solutions, other metric perturbations are induced at order $\epsilon^2$ or higher. In other words, other ten-dimensional perturbations can be induced by expanding the field equations to quadratic order in $\epsilon$ but these perturbations are not present at linear order. Comparing with \eqref{pi-ki} we find that
\be
\epsilon s^I = \frac{a_k^I}{4 k r^k} \label{sprofile}
\ee
which indeed satisfies \eqref{eom-s}. 

For later use, let us note that if $s^I$ depends only on the radial coordinate, $\rho$, then 
\be
D_{\rho} D_{\rho} s^I = (\partial_{\rho}^2 s^I + \frac{1}{\rho} \partial_{\rho} s^I) \qquad
D_{\mu} D_{\nu} s^I = - \frac{1}{\rho} \eta_{\mu \nu} \partial_{\rho} s^I
\ee
are the only non-vanishing components of $D_{m} D_{n} s^I$. Moreover, one can show that 
\be
\rho^2 ( D_{\rho} D_{\rho} s^I +  \eta^{\mu \nu} D_{\mu} D_{\nu} s^I) = \rho^2 \partial_{\rho}^2 s^I - 3 \rho \partial_{\rho} s^I 
=  k (k-4) s^I
\ee
for onshell $s^I$ depending only on the radial coordinate.

The general map between five-dimensional fields (and equations of motion) and ten-dimensional fluctuations was worked out to quadratic order in $\epsilon$ in \cite{Skenderis:2006uy}. 
In particular, working in de Donder-Lorentz gauge, the map between the five-dimensional Einstein metric perturbation $H_{mn}$ and ten-dimensional fields to quadratic order is
\be
H_{mn} = h^0_{mn} + \frac{1}{3} \pi^0 g^o_{mn} + \epsilon^2 \sigma_{(2) mn} \label{kk-redef}
\ee
where $h^0_{mn}$ is the ten-dimensional metric perturbation associated with the trivial harmonic (to order $\epsilon^2$), $\pi^0$ is the trace of the metric perturbation on the $S^5$ associated with the trivial harmonic (to order $\epsilon^2$)\footnote{Note that $\pi^0$ vanishes at linear order in the Coulomb branch solutions.}  and 
\bea
\sigma_{(2)mn} &=  &  \sum_{I} z(k) \left ( A_I D_{p} D_{m} s^I D^p D_{n} s^I + B_I s^I D_m D_n s^I   \right . \label{kk-redef2} \\
 && \qquad \qquad  \left . + D_I ( D_p s^I)(D^p s^I) g^o_{mn} + E_I (s^I)^2 g^o_{mn} \right ) \nn
\eea
where the coefficients $(A_I,B_I,D_I,E_I)$ depend on the degree of the harmonic. Explicit values for the coefficients in the case of $k=2$ were given in \cite{Skenderis:2006uy}:
\be
A_2 = - \frac{4}{9}; \qquad
B_2 = \frac{20}{3}; \qquad
D_2 = - \frac{20}{9}; \qquad
E_2 = \frac{64}{9}. \label{coff1}
\ee
Restricting to the fields depending only on the radial coordinate and working onshell we find that 
\bea
\sigma_{(2) \rho \rho} &=& \sum_I z(k) \left [ (16 A_I + D_I) (\partial_\rho s^I)^2 + (8k (k-4) A_I + 4 B_I) \frac{s^I}{\rho} \partial_{\rho} s^I  \right . \nonumber \\ 
&& \qquad \qquad  \left . + ( E_I + k (k-4) ( k (k-4) A_I + B_I)) \frac{ (s^I)^2}{\rho^2} \right ] \label{redef1} \\ 
\sigma_{(2) \mu \nu} &=& \delta_{\mu \nu}  \sum_I z(k) \left [ (A_I + D_I) (\partial_\rho s^I)^2 -  B_I  \frac{s^I}{\rho} \partial_{\rho} s^I + E_I \frac{ (s^I)^2}{\rho^2} \right ] \nonumber 
\eea
We should note however that the field redefinition \eqref{kk-redef2} gives the reduced metric in a specific gauge: we can always make a diffeomorphism $\xi_{n}$ which is quadratic in $s^I$ such that 
\be
\delta H_{mn} = D_m \xi_n + D_n \xi_m. 
\ee
In the case of interest, such a diffeomorphism must respect the Poincar\'{e} invariance and hence 
\be
\xi_{m} = D_{m} \left ( \sum_{I} z(k) F_I (s^I)^2 \right ) \label{f-def}
\ee
with $F_I$ being arbitrary. The effect of such a diffeomorphism is to shift the coefficients arising in \eqref{redef1}, but the form of the expression remains unchanged. (A natural way to fix the gauge would be to impose a Fefferman-Graham gauge on the resulting five-dimensional metric but this condition was not imposed in \cite{Skenderis:2006uy}). 

We also know from \cite{Skenderis:2006uy} that we can express the terms to quadratic order in $\pi^0$, which we denote as $\pi^0_{(2)}$ as
\be
\pi^0_{(2)} = \epsilon^2 \sum_I z(k) \left ( J_I   ( s^I)^2 + L_I (D_m s^I)(D^m s^I) \right ) \label{pi02}
\ee
where the coefficients $(J_I,L_I)$ can be determined explicitly from the ten-dimensional field equations at quadratic order.  
For $k=2$ these coefficients are 
\be
J_2 = - 72
\qquad L_2 = 8. 
\ee
Note that the coefficients $(A_I,B_I,D_I,E_I,J_I,L_I)$ were not calculated for general values of $k$ in \cite{Skenderis:2006uy}. 

\subsection{Entanglement entropy} \label{section52}

Consider a solution which can be expressed as a perturbation of $AdS_5 \times S^5$ and which preserves full Poincar\'{e} invariance of the dual field theory. Then the metric can be written as 
\be
ds^2 = (g^{o}_{mn} + h_{mn}) dx^{m} dx^{m} + (g^o_{ab} + h_{ab}) dy^a dy^b
\ee
where the metric perturbations depend only on the radial coordinate $\rho$ and on the sphere coordinates $y^a$. Now consider the entanglement entropy for a slab region in the dual field theory, with the slab being defined as the region $- l < x < l$; the slab is assumed to be longitudinal to the the $y$ and $z$ directions.   

We can compute the entanglement entropy from the ten-dimensional metric by finding an eight-dimensional minimal surface on a fixed time slice for which the boundary conditions are $x \rightarrow \pm l$ as $\rho \rightarrow 0$, with the surface wrapping the whole five sphere. From symmetry the minimal surface is specified by the function
\be
x (\rho, y^a).
\ee
We can equivalently express the minimal surface as $\rho(x,y^a)$.
Moreover, working with the leading order metric (which depends only on $\rho$) this function is clearly independent of the spherical coordinates. Thus the entangling surface in the perturbed background can be expressed as 
\be
x(\rho,y^a) = x_0(\rho) + x_1(\rho,y^a) + \cdots 
\ee
where implicitly $x_1(\rho,y^a)$ is linear in the metric perturbations and the ellipses denote higher order corrections. 

The induced metric on the entangling surface is 
\be
\gamma_{\alpha \beta} = g_{AB} \partial_{\alpha} x^{A} \partial_{\beta} x^{B}
\ee
With the static gauge fixing used above the induced metric is therefore
\bea
\gamma_{ij} &=& g_{ij} + g_{xx} \partial_{i}x \partial_j x \\
\gamma_{i a} &=& g_{xx} \partial_i x \partial_a x \nn \\
\gamma_{ab} &=& g_{ab} + \partial_a x \partial_b x \nn
\eea
where $x^i = (\rho,y,z)$ are the non-compact coordinates of the entangling surface. Imposing the further condition that $x$ is independent of $y$ and $z$ we find that 
\be
\gamma_{yy} = g_{yy} \qquad
\gamma_{xx} = g_{xx} \qquad
\gamma_{\rho \rho} = g_{\rho \rho} + g_{xx} (\partial_{\rho} x)^2 \qquad \gamma_{\rho a} = g_{xx} \partial_{\rho} x \partial_a x
\ee
and therefore the determinant of the induced metric is given by 
\be
\sqrt{\gamma} = \sqrt{g_{yy} g_{zz} g_{\rho \rho}} \sqrt{ { \rm{det}} \left ( g_{ab} (1 + \frac{g_{xx}}{g_{\rho \rho}} (\partial_{\rho} x)^2 ) + g_{xx} \partial_a x \partial_b x \right ) }. \label{5dlinear}
\ee
The entanglement entropy functional is then
\be
S = \frac{1}{4 G_{10}} \int d^{3} x d^5 y \sqrt{\gamma}.
\ee

\subsubsection{Linear order}

For an entangling surface lying near the conformal boundary, so that the metric can be expressed as a perturbation of $AdS_5 \times S^5$, the leading contribution to the entanglement entropy is that of a surface in $AdS_5 \times S^5$. Now consider the contribution to the entanglement entropy to linear order in the metric perturbations. Since $x$ is independent of the spherical coordinates $y^a$ to leading order, the term $\partial_a x \partial_b x$ appearing in (\ref{5dlinear}) is at least quadratic in the metric perturbation and can be neglected at this order, so the entanglement entropy is simply:
\be
S = \frac{1}{4 G_{10}} \int d^{3} x d^5 y  \sqrt{g_{yy} g_{zz} (g_{\rho \rho} + g_{xx} (\partial_{\rho} x)^2 )} \sqrt{ { \rm{det}} ( g_{ab})}, \label{ee10}
\ee
where implicitly we work only to linear order in the metric perturbations. However, since we integrate over the five sphere only zero mode spherical harmonics can contribute at linear order and therefore we can substitute
\be
g_{mn} = g^{o}_{mn} + h^0_{mn}; \qquad g_{ab} = g^{o}_{ab} (1 + \frac{1}{5} \pi^0).
\ee
Moreover, we can also express the embedding function in terms of scalar spherical harmonics:
\be
x(\rho,y^a) = x_0(\rho) + \sum_I x^I_1(\rho) Y^I(y^a) + \cdots 
\ee
and again only the zero mode can contribute at this order. Let us denote
\be
\bar{x}(\rho) = x_0(\rho) + x^0_1(\rho), \label{embedinglinear}
\ee
i.e. the embedding function is to this order only dependent on the radial coordinate $\rho$. 

The entanglement entropy integral then factorises as
\bea
S &=& \frac{1}{4 G_{10}} \int d^5 y \sqrt{ {\rm{det}} g^o_{ab}} \int d^3x  \sqrt{g_{yy} g_{zz} (g_{\rho \rho} + g_{xx} (\partial_{\rho} \bar{x})^2 )} (1 + \frac{1}{2} \pi^0) \\
&=& \frac{1}{4 G_5}  \int d^3x  \sqrt{g_{yy} g_{zz} (g_{\rho \rho} + g_{xx} (\partial_{\rho} \bar{x})^2 )} (1 + \frac{1}{2} \pi^0), \nn
\eea
where we use
\be
\frac{1}{G_{10}} = \frac{V_{S^5}}{G_{5}}
\ee
and $V_{S^5}$ is the volume of the five sphere. 

Now let us compare to the entanglement entropy computed directly from the five-dimensional Einstein metric $g^5_{mn}$. This is very similar to the expression above:
\be
S = \frac{1}{4 G_5}  \int d^3x  \sqrt{g^5_{yy} g^5_{zz} (g^5_{\rho \rho} + g^5_{xx} (\partial_{\rho} \bar{x})^2 )}. \label{ee5}
\ee
If we now recall that (up to linear order)
\be
g^{5}_{mn}  =   g^o_{mn} + h^0_{mn} + \frac{\pi^0}{3} g^{o}_{mn}
\ee
we find that the ten-dimensional and five-dimensional expressions precisely agree. 

\bigskip

An alternative derivation of this result can be given using the fact that the change in the entanglement entropy is 
\be
\delta S =  \frac{1}{8 G_{10}} \int d^3 x d^5 y \sqrt{\gamma^o} T^{AB} h_{AB}
\ee
where $T_{AB}$ is the energy momentum tensor of the original minimal surface (with induced metric $\gamma^o$) and $h_{AB}$ is the change in the background metric. 
Using the explicit form of the energy momentum tensor we then find that 
\be
\delta S = \frac{1}{8 G_{10}} \int d^3 x d^5 y \sqrt{\gamma^o} \left ( g^{o ab} h_{ab} + g^{o yy} h_{yy} + g^{o zz} h_{zz} + \gamma^{o \rho \rho} (h_{\rho \rho} + h_{xx} (\partial_{\rho} \bar{x})^2 ) \right ). 
\ee
As above the integration over the five sphere picks out the zero modes in the harmonic expansions of the metric perturbations, resulting in 
\bea
\delta S &=& \frac{1}{8 G_{5}} \int d^3 x \sqrt{\gamma^o} \left ( \pi^0 + g^{o yy} h^0_{yy} + g^{o zz} h^0_{zz} + \gamma^{o \rho \rho} (h^0_{\rho \rho} + h^0_{xx} (\partial_{\rho} \bar{x})^2 ) \right ) \\
&=& \frac{1}{8 G_{5}} \int d^3 x \sqrt{\gamma^o}  \left ( g^{o yy} H_{yy} + g^{o zz} H_{zz} + \gamma^{o \rho \rho} (H_{\rho \rho} + H_{xx} (\partial_{\rho} \bar{x})^2 ) \right ) \nn
\eea
where $H_{mn}$ is the five-dimensional Einstein metric perturbation to linear order, see (\ref{linear-ein}). The latter expression is exactly equivalent to 
\be
\delta S = \frac{1}{8 G_{5}} \int d^3x \sqrt{\gamma^o} T^{mn} H_{mn}
\ee
where $T^{mn}$ is the energy momentum tensor of the minimal surface in five-dimensional anti-de Sitter, thus demonstrating the equivalence between the five-dimensional and ten-dimensional computations.  

\subsubsection{Quadratic order}

Now let us consider the entanglement entropy to quadratic order in the metric perturbations. Since the embedding function is independent of the sphere coordinates to at least quadratic order (c.f. (\ref{embedinglinear})), the expression (\ref{ee10}) is still valid. Moreover, if we expand the embedding as 
\be
x(\rho,y^a) = x_0(\rho) + \epsilon x^0_1(\rho) + \epsilon^2 \sum_I x^{I}_{2}(\rho) Y^I(y) + \cdots 
\ee
we can see that again only the zero mode of the second order term can contribute after integration over the five sphere. Thus $x$ is also independent of the sphere coordinates to this order, and using recursion we see that $x$ depends only on the radial coordinates to {\it all} orders in the expansion. 

Thus the entanglement entropy computed from ten dimensions is
\bea
S &=& \frac{1}{4 G_{10}} \int d^{3} x d^5 y  \sqrt{g_{yy} g_{zz} (g_{\rho \rho} + g_{xx} (\partial_{\rho} x)^2 )} \sqrt{ { \rm{det}} ( g_{ab})}, \label{eeb10} \\
&\equiv&  \frac{1}{4 G_{10}} \int d^{3} x d^5 y \sqrt{ {\rm{det}} \gamma_{ij}}  \sqrt{ { \rm{det}} ( g_{ab})}, \nn
\eea
where $\gamma_{ij}$ is the non-compact part of the induced metric and implicitly $x$ is now taken to depend only on $\rho$. 

To show the equivalence between (\ref{eeb10}) and (\ref{ee5}) we need to know the explicit map between the five-dimensional Einstein metric and the ten-dimensional metric fluctuations to quadratic order. Since this map is not known in full generality, we will focus on the case of general Coulomb branch solutions, using the expressions for perturbations given in section \ref{gcb}. 

We can use the standard identities for expanding determinants to write
\be
\sqrt{ { \rm{det}} ( g_{ab})} =  \sqrt{ { \rm{det}} ( g^o_{ab})} \left ( 1 + \frac{1}{2} h^{a}_{a} + \frac{1}{8} (h^a_a)^2 - \frac{1}{4} h^{ab} h_{ab} + \cdots  \right )
\ee
where 
\be
h^{ab} = g^{o ac} g^{o bd} h_{bd}.
\ee
Now using the expressions given in section \ref{gcb}
\be
h^{a}_{a} = \epsilon \sum_I  (10 k s^I) Y^I + \epsilon^2 \sum_I \pi^I_{(2)} Y^{I} + \cdots,
\ee
where we will need only the constant harmonic term at quadratic order, $\pi^0_{(2)}$, which is given in \eqref{pi02}. 
Similarly
\be
h^{ab} h_{ab} = 20 \epsilon^2 \sum_{I,J} (k_I s^I Y^I)(k_J s^J Y^J) + {\cal O}(\epsilon^3)
\ee
and thus to order $\epsilon^2$
\be
\sqrt{ { \rm{det}} ( g_{ab})} =  \sqrt{ { \rm{det}} ( g^o_{ab})} \left ( 1 + 5  \epsilon \sum_I k s^I Y^I  + \frac{1}{2} \epsilon^2 \pi^0_{(2)} + \frac{15}{2} \epsilon^2 
 \sum_{I,J} k_I s^I  k_J s^J Y^I Y^J +  \cdots  \right ). \label{sph}
\ee
Here the ellipses denote terms of $\epsilon^3$ and higher, as well as terms at order $\epsilon^2$ which are linear in spherical harmonics (and hence integrate to zero over the five sphere). 

The non-compact components of the metric can be expressed as 
\be
g_{mn} = g^{o}_{mn} + \epsilon \sum_I h^I_{(1) mn} Y^I + \epsilon^2 \sum_I h^{I}_{(2) mn} Y^I + \cdots \label{non-compact}
\ee
where $h^{I}_{(2) mn}$ is quadratic in $s$. The explicit form can be determined by the ten-dimensional supergravity equations at quadratic order in $\epsilon$, see \cite{Skenderis:2006uy}, but will not be needed here. The non-compact part
of the induced metric inherits an analogous expansion in powers of $\epsilon$:
\be
\gamma_{ij} = \gamma^{o}_{ij} + \epsilon \gamma_{(1) ij} + \epsilon^2 \gamma_{(2) ij}  + \cdots
\ee
where
\bea
\gamma^{o}_{ij} &=& g^{o}_{ij} + g^{o}_{xx} \partial_i x^o \partial_j x^o; \\
\gamma_{(1) ij} &=&   \left ( \sum_I (h^I_{(1) ij} + h^I_{(1) xx} (\partial_i x^o) (\partial_j x^o) ) Y^I  \right ) + g^{o}_{xx} (\partial_i x^o \partial_j x_{(1)} + \partial_i x_{(1)} \partial_j x^o); \nn \\
& \equiv & \sum_I \gamma_{(1) ij}^I Y^I; \nn \\
\gamma_{(2) ij} &=&   \left ( \sum_I (h^I_{(2) ij} + h^I_{(2) xx} (\partial_i x^o) (\partial_j x^o) ) Y^I  \right ) + g^{o}_{xx} (\partial_i x^o \partial_j x_{(2)} + \partial_i x_{(2)} \partial_j x^o)\nn \\
&& +  g^{o}_{xx} \partial_i x_{(1)} \partial_j x_{(1)}; \nn \\
& \equiv & \sum_I \gamma^I_{(2)ij} Y^I. 
\eea
(In the case of interest we have already shown that the embedding function depends only on the $\rho$ coordinate but we write the above expressions more generally.)

Expanding out the induced metric determinant then gives
\be
\sqrt{ { \rm{det}} ( \gamma_{ij})} =  \sqrt{ { \rm{det}} ( \gamma^o_{ij})} \left ( 1 + \frac{1}{2} (\epsilon \gamma^i_{(1) i} + \epsilon^2 \gamma^i_{(2)i}) 
+ \frac{\epsilon^2}{8} (\gamma_{(1)i}^i)^2 - \frac{\epsilon^2}{4} \gamma_{(1)}^{ij} \gamma_{(1)ij} + \cdots  \right ) \label{ncep}
\ee
where
\be
\gamma_{(1)}^{ij} = \gamma^{o ik} \gamma^{o jl} \gamma_{(1) kl} \qquad
\gamma_{(2)}^{ij} = \gamma^{o ik} \gamma^{o jl} \gamma_{(2) kl}.
\ee
Substituting (\ref{sph}) and (\ref{ncep}) into (\ref{eeb10}) and integrating over the five-sphere we then obtain to linear order in $\epsilon$
\be
S = \frac{1}{4 G_5} \int d^3 x \sqrt{ { \rm{det}} ( \gamma^o_{ij})} \left (1 + \epsilon \gamma^{o ij} \partial_i x^o \partial_j x_{(1)} \right ),
\ee
i.e. all terms linear in metric perturbations vanish since they are associated with degree $k \ge 2$ spherical harmonics which integrate to zero over the sphere. Since the 
five-dimensional Einstein metric is unchanged to this order, the entangling surface is also unchanged i.e. $x_{(1)} = 0$.

Dropping terms involving $x_{(1)}$, the contributions to the entanglement entropy functional at order $\epsilon^2$ are 
\bea
\delta S &=& \frac{\epsilon^2 }{4 G_5} \int d^3 x \sqrt{ {\rm {det}} (\gamma^o)}  \left ( \frac{5}{2} \sum_I k z(k) s^I ( \gamma^{I i}_{(1) i} + 3 k s^I) 
+ \frac{1}{8} \sum_I z(k) (\gamma^{Ii}_{(1) i})^2 \right .  \label{110d} \\
&& \qquad \qquad \qquad \left . - \frac{1}{4} \sum_I z(k) \gamma^{I ij}_{(1)} \gamma^I_{(1) ij} + \frac{1}{2} \gamma^{0i}_{(2) i} + \frac{1}{2} \pi^0_{(2)}  \right ) \nn
\eea
where we define $z(k)$ as the spherical harmonic normalisation 
\be
\int d^5y \sqrt{{\rm {det}} g_{ab}^o} Y^I Y^J = z(k) \delta^{IJ} V_{S^5}.
\ee
The corresponding expression for the contribution to the five-dimensional entanglement entropy at quadratic order is: 
\be
\delta S = \frac{\epsilon^2}{4 G_5} \int d^3 x \sqrt{ {\rm {det}} (\gamma^o)}  \left ( \frac{1}{2} H^i_{(2)i} + \frac{1}{2} H_{(2) xx} (\partial^i x^o)(\partial_i x^o) + \partial^i x^o \partial_i x_{(2)} \right ) \label{15d}
\ee
where $H_{(2) mn}$ is the quadratic correction to the five-dimensional Einstein metric and implicitly indices are raised with $\gamma^{o ij}$. Let us split $H_{(2)mn}$ as 
\be
H_{(2) mn} = h^0_{(2) mn} + \frac{1}{3} \pi_{(2)}^0 g^o_{mn} + \sigma_{(2) mn}
\ee 
where $\sigma_{(2) mn}$ defines the field redefinition and is quadratic in $s$, while $\pi_{(2)}^0$ is also quadratic in $s$.  

To match (\ref{110d}) and (\ref{15d}) one requires that 
\bea
 &&\sigma^i_{(2)i} + \sigma_{(2) xx} (\partial^i x^o)(\partial_i x^o)   \label{general} \\
&& \qquad = 5 \sum_I k z(k) s^I ( \gamma^{I i}_{(1) i} + 3 k s^I) + \frac{1}{4} \sum_I z(k) (\gamma^{Ii}_{(1) i})^2
- \frac{1}{2} \sum_I z(k) \gamma^{I ij}_{(1)} \gamma^I_{(1) ij}  \nn
\eea
To interpret this relationship, it is useful to consider first the case of an infinite strip. For an infinite strip, the entangling surface in $AdS$ is described by constant $x^o$ and extends throughout the bulk. In this case the entangling surface extends beyond the asymptotic region in which the geometry can be expressed as a perturbation of $AdS_5 \times S^5$, but nonetheless one can match the integrands for the five-dimensional and ten-dimensional entanglement entropy in the asymptotic region by setting $x^o$ to be constant in \eqref{general}. 

In the case of an infinite strip $x^o$ is constant and dropping these terms gives
\be
 \sigma^i_{(2)i}  =  \sum_I  z(k) \left ( 5 k s^I (h^{I i}_{(1) i} + 3 k s^I)  + \frac{1}{4}  (h^{Ii}_{(1) i})^2
- \frac{1}{2}  h^{I ij}_{(1)} h^I_{(1) ij} \right ),
\ee
where now indices are raised by $\gamma^{oij} \equiv g^{o ij}$, i.e. the hyperbolic metric. This expression reduces to
\be
\sigma_{(2)i}^i   =   \sum_{I} \frac{ 16 z(k)}{(k+1)^2}  \left (- 8 \rho^2 (\partial_\rho s^I)^2 + k (15-k) \rho s^I \partial_{\rho} s^I + k^2 (k - 7) (s^I)^2  \right ) \label{final}
\ee
However, using \eqref{redef1} in  section \ref{gcb}, one can show that for perturbations $s^I$ which depend only on the radial coordinate
\bea
\sigma^i_{(2)i} &=& \sum_I z(k) \left [ (18  A_I + 3 D_I)  \rho^2(\partial_\rho s^I)^2 + (8k (k-4) A_I +  2 B_I) \rho s^I \partial_{\rho} s^I  \right . \nonumber \\ 
&& \qquad \qquad  \left . + ( 3 E_I + k (k-4) (k (k-4) A_I +  B_I))(s^I)^2 \right ], \label{KK}
\eea
where the coefficients $(A_I,B_I,D_I,E_I)$ for $k=2$ are given in \eqref{coff1}. Taking into account an appropriate choice of diffeomorphism $F_I$, defined in \eqref{f-def}, this indeed matches \eqref{final}. 

For general $k$ the coefficients $(A_I,B_I,D_I,E_I)$ were not computed in \cite{{Skenderis:2006uy}}. Nonetheless, it is apparent that \eqref{KK}
has the same structure as \eqref{final} and we can argue as follows that these expressions agree, mode by mode, even without knowing the explicit expressions for the coefficients. We have already shown that the ten-dimensional and five-dimensional entanglement entropies agree for Coulomb branch solutions which admit consistent truncations. For such solutions  (\ref{final}) agrees with the result that one gets from direction reduction (\ref{KK}). Moreover, the matching between (\ref{final}) and (\ref{KK}) arises mode by mode, as the fields $s^I$ associated with spherical harmonics of different rank $k$ have different functional dependence on the radial coordinate, see \eqref{sprofile}, and cannot cancel each other. 

Since the agreement between (\ref{final}) and (\ref{KK}) holds for all consistent truncations with different symmetry groups and different profiles for the scalar fields, this implies that the coefficients of the terms $(\partial_{\rho} s^I)^2$, $s^I \partial_{\rho} s^I$ and $(s_I)^2$ must match between the left and right hand sides of (\ref{final}). However, since these coefficients match for all solutions with consistent truncations, they also match for solutions which do not admit consistent truncations and therefore the matching of five-dimensional and ten-dimensional entanglement entropy holds for entangling surfaces in all Coulomb branch solutions, up to quadratic order in the expansion parameter. The same argument can be used for  strip entangling regions, i.e. (\ref{general}), and indeed for generic shape entangling regions. 

\subsection{Summary and interpretation}

Let us summarise what has been proven in this section. We considered solutions of ten-dimensional type IIB supergravity which respect the Poincar\'{e} invariance of the dual field theory; the Einstein frame metric in ten dimensions is therefore of the form
\be
ds^2 = g_{\rho \rho} (\rho,y^a) d \rho^2 + g_{\mu \nu}(\rho,y^a) dx^{\mu} dx^{\nu} + g_{ab}(\rho,y^a) dy^a dy^b,
\ee
where we choose a gauge in which $g_{\rho a} = 0$. We also assumed that the geometry is asymptotic to $AdS_5 \times S^5$ so that as $\rho \rightarrow 0$
\be
g_{\rho \rho} \rightarrow  \frac{1}{\rho^2} \qquad
g_{\mu \nu} \rightarrow \frac{1}{\rho^2} \eta_{\mu \nu} \qquad
g_{ab} \rightarrow g^o_{ab} 
\ee
where $g^o_{ab}$ is the metric on the unit $S^5$. 

We then computed the entanglement entropy for a strip region in the dual field theory by finding a codimension two minimal surface on a surface of constant time which asymptotically wraps the five sphere. Working in the region near the conformal boundary in which all metric coefficients can be expanded perturbatively in a Fefferman-Graham expansion in the basis of spherical harmonics, we showed that such an entangling surface depends only on the radial coordinate $\rho$ to all perturbative orders, i.e. it is described by $x(\rho)$ with the width of the strip being $l$ on the conformal boundary $\rho \rightarrow 0$. 

\begin{figure}
\begin{center}
\setlength{\unitlength}{0.50mm}
\includegraphics*[width=0.8\linewidth]{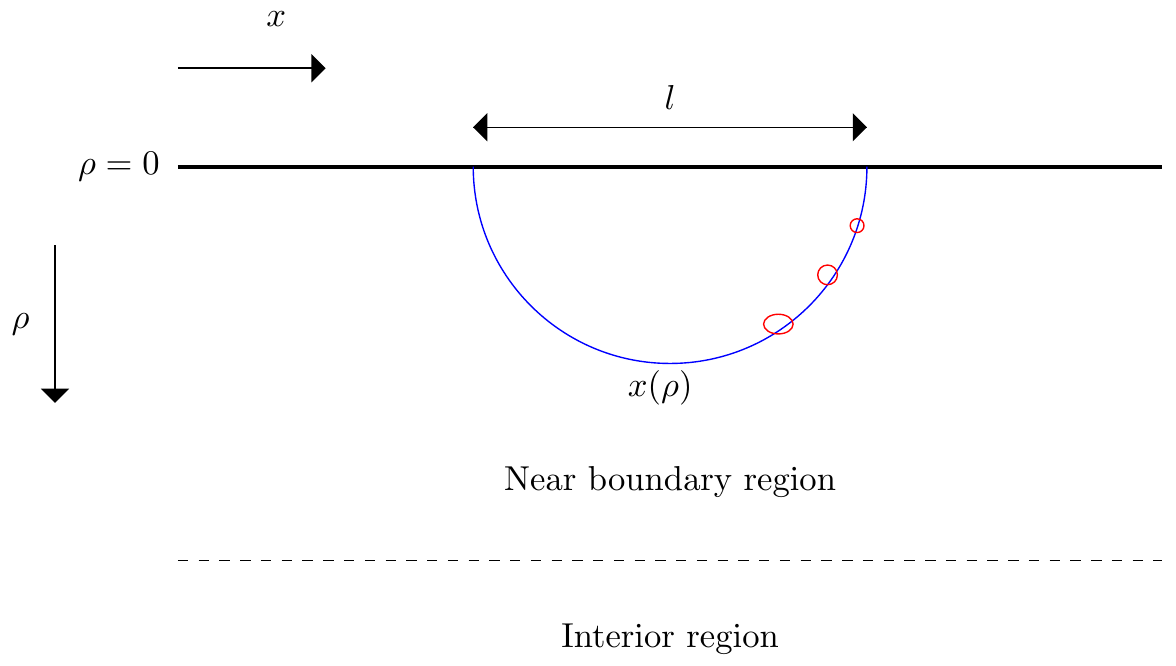}
\caption{We consider entangling surfaces which are contained within the near boundary region where the Fefferman-Graham expansion of the metric may be used. At each point on the three-dimensional Ryu-Takayanagi minimal surface (shown in blue), there is a five-dimensional compact space (shown in red) which is topologically a five sphere.}
\label{fig:slabsurface}
\end{center}
\end{figure}

As an immediate consequence of the minimal surface being described by $x(\rho)$, the induced metric $\gamma_{\alpha \beta}$ on the minimal surface factorises:
\be
\gamma_{\rho \rho} = g_{\rho \rho} + g_{xx} (x')^2 \qquad
\gamma_{yy} = g_{yy} \qquad \gamma_{zz} = g_{zz} \qquad \gamma_{ab} = g_{ab}.
\ee
The eight-dimensional minimal surface is therefore topologically a product of a three-dimensional surface and a five-sphere, see Figure~\ref{fig:slabsurface}.
It is nonetheless non-trivial to show that the area of this minimal surface gives the entanglement entropy computed from the five-dimensional perspective. 

The induced metric depends explicitly on both the radial coordinate and the spherical coordinates, so one cannot trivially integrate over the spherical coordinates. In addition, the relationship between the five-dimensional Einstein metric $g^{5}_{mn}$ occurring in the Ryu-Takayanagi formula and the ten-dimensional Einstein metric is extremely complicated, involving derivative field redefinitions. 

The induced metric on the co-dimensional two Ryu-Takayanagi surface is 
\be
\gamma^5_{\rho \rho} = g^5_{\rho \rho} + g^5_{xx} (x')^2 \qquad
\gamma^5_{yy} = g^5_{yy} \qquad \gamma^5_{zz} = g^5_{zz}. 
\ee
Working up to quadratic order in the perturbation relative to $AdS \times S^5$, i.e. to order $(h_{AB})^2$ in $g_{AB} = g^{o}_{AB} + h_{AB}$, we showed that the ten-dimensional entanglement entropy 
\be
S = \frac{1}{4 G_{10}} \int d^3 x d^5 y \sqrt{\gamma}
\ee
indeed agrees with the five-dimensional Ryu-Takayanagi computation
\be
S = \frac{1}{4 G_{5}} \int d^3 x \sqrt{\gamma^5}
\ee
when we take into account the reduction map. More precisely, the Ryu-Takayanagi integrand matches the top-down integrand once the latter is integrated over the five-sphere: the volume form of the Ryu-Takayanagi minimal surface matches the volume form of the top-down minimal surface, once the latter is integrated over the spherical coordinates. 

\bigskip

Before leaving this section, we should mention another related  test of the top-down entanglement entropy formula using Kaluza-Klein holography. Entanglement entropy for asymptotically $AdS_3 \times S^3$ geometries corresponding to 1/4 and 1/8 BPS geometries associated with black hole microstates was computed in \cite{Giusto:2015dfa}. The entanglement entropy was computed using both the top down prescription, i.e. codimension two minimal surfaces in six dimensions, and by applying the Ryu-Takayanagi formula to the three-dimensional Einstein metric extracted using Kaluza-Klein holography. The results were in agreement, working up to quadratic order in perturbations around $AdS_3 \times S^3$, as in the asymptotically $AdS_5 \times S^5$ case analysed above.

\section{Unquenched Flavor Solutions} \label{six}

Another example for studying top down entanglement entropy is provided by unquenched flavor solutions, i.e. systems of flavor branes in which the backreaction of the branes onto the metric has been computed, working perturbatively in the ratio of flavors to colors. 

The computation of the backreaction is most tractable when the branes are smeared over transverse directions. In particular, \cite{Bigazzi:2009bk, Nunez:2010sf} discuss the case of the massless D3/D7 system in which probe D7 branes are smeared over the transverse $S^2$ space. The system is type IIB supergravity coupled to D7-brane sources, and \cite{Bigazzi:2009bk, Nunez:2010sf} takes the following ansatz for the Einstein frame metric in the supersymmetric (zero temperature) case:
\begin{equation}
ds^2_{10}=h^{-\frac{1}{2}} dx_{\mu} dx^{\mu}+h^{\frac{1}{2}}\left[F^2 d\varrho^2+S^2 ds^2_{KE}+F^2(d\tau +A_{KE})^2 \right]
\end{equation}
where the functions $h(\varrho)$, $S(\varrho)$, $F(\rho)$ only depend on the radial coordinate $\varrho$, and the five-dimensional Sasaki-Einstein manifold $X^5$ is written as a $U(1)$ fibration over a 4d K\"ahler-Einstein base. For $X=S^5$ the KE base is $CP^2$. 

To compute the entanglement entropy for a strip in the $x$-direction from 10d we follow the usual procedure. By symmetry the embedding is given by $x = x(\varrho)$
and the induced metric on the embedding surface is:
\begin{equation}
ds^2_{8}=h^{-\frac{1}{2}} \left[dy^2+dz^2 +x'^2d\varrho^2 \right]+h^{\frac{1}{2}}\left[ F^2 d\varrho^2+S^2 ds^2_{KE}+F^2(d\tau +A_{KE})^2 \right],
\end{equation}
where $x' \equiv dx/d\varrho$. Note that the minimal surface wraps the entire internal space. 

One thus finds that the entanglement entropy functional is:
\be
S = \frac{1}{4 G_{10}} \int d^3 x d^5 y   \sqrt{\textrm{det}\gamma_8}
\ee
where 
\begin{equation}
\sqrt{\textrm{det}\gamma_8}=h^{\frac{1}{2}}  F S^4  \sqrt{x'^2+h  F^2}  \sqrt{\textrm{det}g_{X^5}}.
\end{equation}
Since this determinant factorises we can immediately integrate over the internal space to obtain
\be
S = \frac{V_{X^5}}{4 G_{10}} \int d^3 x  h^{\frac{1}{2}}  F S^4  \sqrt{x'^2+h F^2} \label{10dquench}
\ee
where the integration is over $(\varrho,y,z)$ and we define
\be
 V_{X^5} = \int d^5 y  \sqrt{\textrm{det}g_{X^5}}.
\ee

Explicit expressions for the metric functions were calculated in \cite{Bigazzi:2009bk, Nunez:2010sf} working perturbatively in the number of flavors:
\bea
S &=& \alpha'^{\frac{1}{2}}e^{\varrho}\left(1+\epsilon_{*}\left(1/6+\varrho_*-\varrho \right) \right)^{\frac{1}{6}}; \\
F &=& \alpha'^{\frac{1}{2}}e^{\varrho}\left(1+ \epsilon_{*}(\varrho_*-\varrho)\right)^{\frac{1}{2}}\left(1+\epsilon_{*}\left(1/6+\varrho_*-\varrho \right) \right)^{-\frac{1}{3}}; \nn \\
\frac{dh}{d\varrho} & =& -Q_c \alpha'^{-2}e^{-4\varrho}\left(1+\epsilon_{*}\left(1/6+\varrho_*-\varrho \right) \right)^{-\frac{2}{3}}. \nn
\eea
Here $\varrho_*$ is a scale that is introduced for convenience; $Q_c$ is proportional to the number of colors and $\epsilon_*\equiv Q_f e^{\Phi_*}$ is the small expansion parameter: $Q_f$ is proportional to the number of flavors and $\Phi_*$ is the value of the dilaton at $\varrho_*$. 

The equation for $dh/d\varrho$ can be integrated up to express $h(\varrho)$ in terms of incomplete gamma functions, with integration constant being determined by the requirement that the metric is asymptotically anti-de Sitter. 
Using the explicit expressions above and expanding in $\epsilon_*$ one finds:
\begin{equation}
h^{\frac{5}{4}}S^4 F=\frac{Q_c^{\frac{5}{4}}}{4\sqrt{2}}+\frac{Q_c^{\frac{5}{4}}}{32\sqrt{2}}\epsilon_*+\frac{Q_c^{\frac{5}{4}}(19+48\varrho -48\varrho_*)}{1536\sqrt{2}}\epsilon_*^2+\mathcal{O}(\epsilon_*^3)
 \end{equation} 
Note that to order $\epsilon_*$ this expression is independent of the radial coordinate. 

\bigskip

Now let us turn to the calculation of the entanglement entropy from the five-dimensional perspective. Let us first note that it is clear that the five-dimensional Einstein
metric cannot be identified as just the non-compact part of the above metric, i.e.
\begin{equation}
ds^2_{5}=h^{-\frac{1}{2}} dx_{\mu} dx^{\mu}+h^{\frac{1}{2}} F^2 d\varrho^2.
\end{equation}
Using the latter metric the computation of the entanglement entropy from five dimensions would be
\be
S = \frac{1}{4 G_5} \int d^3 x \sqrt{\textrm{det}\gamma_3}
\ee
where 
\begin{equation}
\sqrt{\textrm{det}\gamma_3}=h^{-\frac{3}{4}}\sqrt{x'^2+h F^2}.
\end{equation}
This does not agree with the ten-dimensional result;  the latter contains also an additional factor $h^{\frac{5}{4}}S^4 F$ which as we showed above depends on the radial coordinate $\varrho$. 

\subsection{Linear order}

To extract the correct five-dimensional Einstein metric we can again use Kaluza-Klein holography. Working to zeroth order in $\epsilon^{\ast}$ the metric is
\be
ds^2_{10}=  \frac{2 \alpha'}{\sqrt{Q_c}}  e^{2 \varrho} dx_{\mu} dx^{\mu}+ \frac{\sqrt{Q_c}}{2}  d\varrho^2 + \frac{\sqrt{Q_c}}{2} \left [ ds^2_{KE}+ (d\tau +A_{KE})^2 \right].
\ee
By rescaling the coordinates one can pull out an overall factor as 
\be
ds^2_{10} =  \frac{\sqrt{Q_c}}{2} \left [ e^{2 \varrho} d\tilde{x}_{\mu} d\tilde{x}^{\mu}+ d\varrho^2 +  ds^2_{KE}+ (d\tau +A_{KE})^2 \right ]
\ee
where 
\be
\tilde{x}_{\mu} = \frac{2 \sqrt{\alpha'}}{\sqrt{Q_c}} x_{\mu}
\ee
For computational convenience, and to match the conventions of earlier sections, we will set $\sqrt{Q_c}/2 = \sqrt{\alpha'} = 1$; these factors can be reinstated if required. The leading order metric is therefore the produce of $AdS_5$ (in domain wall coordinates) with the Sasaki-Einstein space.

Now let
\be
S = S^o (1+ \delta S); \qquad
F = F^o (1 + \delta F);  \qquad
h = h^o (1  + \delta h),
\ee
where the superscript refers to the value in the $AdS_5 \times S^5$ background and the perturbations are expressed as power series in the parameter $\epsilon_*$. The explicit forms for the perturbations are:
\bea
\delta S &=& \epsilon_* \left ( \frac{1}{36} + \frac{1}{6} (\varrho_* - \varrho) \right ) + \cdots \\
\delta F &=& \epsilon_* \left ( - \frac{1}{18} + \frac{1}{6} (\varrho_* - \varrho) \right ) + \cdots \nn \\
\delta h &=& \epsilon_* \left ( \frac{1}{18} - \frac{2}{3} (\varrho_* - \varrho) \right ) + \cdots \nn
\eea
The metric can as before be written as 
\be
g_{AB} = g^o_{AB} + h_{AB} 
\ee
where 
\bea
h_{\mu \nu} &=& e^{2 \varrho} \left (- \frac{1}{2} \delta h + \frac{3}{8} (\delta h)^2 + \cdots \right ) \eta_{\mu \nu} \label{noncompact1} \\ 
&=& e^{2 \varrho}\epsilon_* \left ( - \frac{1}{36} + \frac{1}{3} (\varrho_{\ast} - \varrho) \right )  \eta_{\mu \nu}+ \cdots   \nn\\
h_{\varrho \varrho} = &=& 2 \delta F + \frac{1}{2} \delta h + \delta F^2 - \frac{1}{8} \delta h^2 + \delta h \delta F \label{noncompact2} \\
&=&  - \frac{1}{12} \epsilon_* + \cdots   \nn 
\eea
while along the compact space
\be
h_{ab} dy^a d y^b = h_{KE} ds_{KE}^2 + h_{\tau \tau} (d \tau + A_{KE})^2
\ee
with 
\be
h_{KE} = 2 \delta S + \frac{1}{2} \delta h + \delta S^2 - \frac{1}{8} \delta h^2 + \delta h \delta S  = \frac{1}{12} \epsilon_* + \cdots   
\ee
and $h_{\tau \tau } = h_{\varrho \varrho}$. Here the ellipses denote terms of order $\epsilon_*^2$ and higher.  

As in previous sections, the metric perturbations can be expressed in the complete basis of harmonics. For the metric perturbations in the non-compact directions,
this expansion involves only the constant harmonic, i.e. 
\be
h_{mn} \equiv h^0_{mn}.
\ee
since (\ref{noncompact1})-(\ref{noncompact2}) are independent of the compact space coordinates. Now consider the perturbations along the compact space. The trace of the metric perturbation
\be
h^{a}_{a} = g^{o ab} h_{ab} = 4 h_{KE} + h_{\tau \tau}
\ee
is independent of the compact space coordinates, and therefore the expansion of the trace in harmonics involves only the constant harmonic 
\be
h^{a}_{a} \equiv \pi^0 = \frac{1}{4} \epsilon_\ast  + \cdots
\ee
We will discuss the decomposition of the traceless part into harmonics below. 

To linear order in the metric perturbations
the correction to the five-dimensional Einstein metric is 
\be
H_{mn} = h^0_{mn} + \frac{1}{3} \pi^0 g^o_{mn}
\ee
and therefore to linear order in $\epsilon_*$
\be
H_{\mu \nu} = \epsilon_{\ast} \left ( \frac{1}{18} + \frac{1}{3} (\varrho_{\ast} - \varrho) \right ) e^{2 \varrho} \eta_{\mu \nu} + \cdots 
\qquad H_{\varrho \varrho} = {\cal O}(\epsilon_{\ast}^2)
\ee
This defines the five-dimensional Einstein metric to linear order. 

We already showed that the entanglement entropy for a strip computed in the ten-dimensional metric $g_{AB} = g^o_{AB} +  h_{AB}$  is always equivalent, to linear order in the perturbations, to the entanglement entropy computed using the five-dimensional Einstein metric $g^5_{mn} = g^o_{mn} + H_{mn}$. This general result implies that (\ref{10dquench}) is indeed equivalent to 
\be
S = \frac{1}{4 G_5} \int d^3 x \sqrt{g^5_{yy} g^5_{zz}} \sqrt{g^5_{\rho \rho} + g^5_{xx} (x')^2} \label{5dquench}
\ee
at linear order in the perturbations. One can show the equivalence directly using the identifications 
\be
g^5_{\mu \nu} = h^{\frac{1}{3}}F^{\frac{2}{3}} S^{\frac{8}{3}} \eta_{\mu \nu} \qquad
g^5_{\rho \rho} = h^{\frac{4}{3}} F^{\frac{8}{3}} S^{\frac{8}{3}}, \label{relations}
\ee
to linear order in $\epsilon_*$. 

\subsection{Non-linear order}

The traceless part of the metric perturbation on the compact space is 
\be
h_{(ab)} dy^a dy^b = \left  ( \frac{h_{KE}}{5} - \frac{h_{\tau \tau}}{5} \right ) ds_{KE}^2 + \left  ( \frac{4 h_{\tau \tau}}{5} - \frac{4 h_{KE}}{5} \right  ) (d \tau + A_{KE})^2
\ee
Working to linear order in the perturbations 
\be
h_{(ab)} dy^a dy^b = \left (\frac{1}{30} \epsilon_* + \cdots  \right ) ds_{KE}^2 +  \left ( - \frac{2}{15} \epsilon_* + \cdots \right  ) (d \tau + A_{KE})^2
\ee
We can now project this onto harmonics:
\be
h_{(ab)} = \sum \left ( \phi^{I_t} Y^{I_t}_{(ab)} + \psi^{I_v} D_{(a} Y^{I_v}_{b )} + \chi^I D_{(a} D_{b)} Y^I \right ).
\ee
For example
\be
\int  D^a h_{(ab)} D^b Y^I d \Omega = 4 \Lambda^I \left ( \frac{\Lambda^I}{5} - 1 \right ) z(k) \chi^I,
\ee
where 
\be
\Box Y^I = \Lambda^I Y^I
\ee
and $d \Omega$ is the volume element on the Sasaki-Einstein, with $z(k)$ the harmonic normalisation. While $h_{(ab)}$ does not depend on the Sasaki-Einstein coordinates, all individual harmonics depend on the coordinates and $h_{(ab)}$ is therefore decomposed into an infinite series of harmonics, as one would have anticipated, given the smearing.

As in section \ref{kk-hol}, the perturbations associated with non-trivial harmonics do not contribute to the entanglement entropy at linear order, but they do contribute at non-linear order. Unfortunately the non-linear relation between the five-dimensional Einstein metric and the ten-dimensional metric is not known for general perturbations in which $\phi^{I_t}$, $\psi^{I_v}$ and $\chi^I$ are non-zero and therefore we cannot check the equivalence of five-dimensional and ten-dimensional entanglement entropy to non-linear order. 

It is interesting to note, however, that the ten-dimensional entanglement entropy (\ref{10dquench}) can  be expressed in five-dimensional form (\ref{5dquench}) provided that one makes the identifications (\ref{relations}). This suggests that the five-dimensional Einstein metric at non-linear order is simply
\be
ds^2 = h^{\frac{1}{3}} F^{\frac{2}{3}} S^{\frac{8}{3}} \left ( h F^2 d \varrho^2 + \eta_{\mu \nu} dx^{\mu} dx^{\nu} \right ).  
\ee
One could explore  whether this is indeed the correct expression for the five-dimensional Einstein metric by checking whether it gives the expected forms for e.g. one point function and higher correlation functions of the holographic stress energy tensor. 

\bigskip

Finally, we should note that very similar analysis should be applicable to smeared solutions in other dimensions including \cite{Bea:2013jxa}; one would need to set up Kaluza-Klein holography for ABJM to explore this case. 

\section{General case} \label{seven}

In this section we consider ten-dimensional asymptotically $AdS_5 \times S^5$ type IIB solutions which respect the Poincar\'{e} invariance of the dual field theory. The ten-dimensional Einstein metric therefore takes the form
\be
ds^2 = g_{\rho \rho} d\rho^2 + g_{\mu \nu} dx^{\mu} dx^{\nu} + 2 g_{\rho a} d\rho dy^a + g_{ab} dy^a dy^b
\ee
where $g_{\mu \nu} \propto \eta_{\mu \nu}$ and all metric components depend on $(\rho,y^a)$. For simplicity let us focus on the case in which the $SO(6)$ symmetry is broken to $SO(5)$, so that the metric depends only on $\rho$ and a single angular coordinate $\theta$. (Examples of such supergravity solutions would be D3-brane Coulomb branch solutions in which all branes lie along a line.)

Suppose we make a coordinate redefinition $(\rho,\theta) \rightarrow (r,\vartheta)$ to bring the metric into the following form
\be
ds^2 = e^{2 B (r,\vartheta)} (dr^2 + e^{2 A (r)} dx^{\mu} dx_{\mu} ) + 2 {\cal A}  (r,\vartheta) dr d \vartheta + g_{\vartheta \vartheta}(r,\vartheta) d \vartheta^2 + g_{S^4} (r, \vartheta) d \Omega_{S^4},
\ee
with  $r$ being the radial coordinate of the five-dimensional metric in Einstein frame:
\be
ds^2_5 = dr^2 + e^{2 A(r)} dx^{\mu}dx_{\mu}.
\ee 
For consistent truncations, we know the explicit form of the map between five and ten-dimensional solutions, i.e. the explicit form of $B(r,\vartheta)$ etc. In the vicinity of the conformal boundary one can use Kaluza-Klein holography to work out the map as a power series in the radial coordinate. 

Deep in the interior of a general such spacetime we do not know the explicit form of the map, but such a map must exist. Note however that the causal structures of the five-dimensional Einstein metric and the ten-dimensional Einstein metric do not necessarily agree: even in consistent truncations, the former can be singular while the latter is smooth.  
The choice of a specific ten-dimensional radial coordinate adapted to the five-dimensional Einstein metric corresponds to identifying the RG scale of the dual field theory. 
The five-dimensional Einstein metric would in general be supported by the stress energy tensor associated with the entire tower of Kaluza-Klein modes. 

Next consider the simplest possible entangling surface in this geometry, corresponding to the half plane entangling region $x > 0$ in the dual field theory. On symmetry grounds, 
the bulk entangling surfaces are codimension two surfaces $x=0$ at constant time. Agreement between the ten-dimensional and five-dimensional entanglement entropies requires 
\be
\frac{1}{4 G_{10}} \int d^3 x d^5 y \sqrt{\gamma} = \frac{1}{4 G_5} \int d^3 x \sqrt{\gamma_5}
\ee
which in turn requires that 
\be
\int dr d \vartheta e^{2 B + 2 A} \left ( e^{2 B} g_{\vartheta \vartheta} - {\cal A}^2 \right )^{\frac{1}{2}} \sqrt{\rm {det} g_{S^4}} = \pi^3 \int dr e^{2 A}. 
\ee
In this paper we have effectively checked that this relation holds in all cases in which we can independently calculate the five-dimensional Einstein metric. In cases where the five-dimensional Einstein metric is not known, one may be able to deduce the five-dimensional Einstein metric by insisting that this expression hold. 

\subsection{Relation to Lewkowycz-Maldacena derivation}

In this section we will explain the origin of the top down entanglement entropy formula, using a similar approach to Lewkowycz-Maldacena in \cite{Lewkowycz:2013nqa}. 

The entropy associated with a given density matrix $\rho$ can be computed using the replica trick as
\be
S = - n \partial_n \left [ \log Z(n) - n \log Z(1) \right ]_{n=1} \label{lm1}
\ee
where 
\be
Z(n) = {\rm Tr}(\rho^n).
\ee
Here $Z(1)$ can be computed by considering (Euclidean) evolution on a circle, i.e.
\be
\rho = {\cal P} \exp \left ( - \int_{\tau_0}^{\tau_0 + 2 \pi} d \tau H(\tau) \right )
\ee
where $H$ is the Hamiltonian and the periodicity of the $\tau$ direction is $2 \pi$. 
$Z(n)$ is then computed by considering the evolution over a circle
 of $n$ times the length of the original circle. 
 
 \begin{figure}
\begin{center}
\setlength{\unitlength}{0.50mm}
\includegraphics*[width=0.7\linewidth]{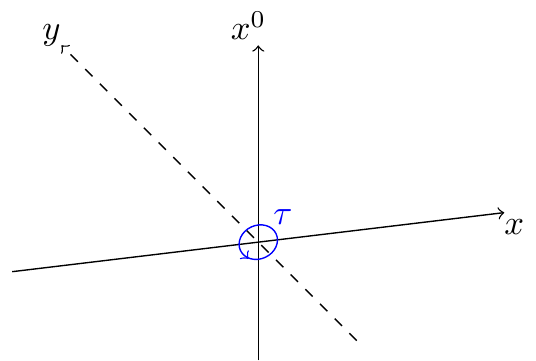}
\caption{The half space entangling region $x \ge 0$, with boundary the y-axis. The coordinate $\tau$ is the polar coordinate in the plane of $x$ and the Euclidean time $x^0$.}
\label{fig:replica1}
\end{center}
\end{figure} 
 
 In the context of thermal density matrices the circle direction is Euclidean time. For entanglement entropy, the appropriate circle direction is that enclosing the boundary of the entangling region, see the example shown in Figure~\ref{fig:replica1}. The well-known CHM map relates certain thermal entropies to entanglement entropies in conformal field theories \cite{Casini:2011kv}.

To compute the entanglement entropy holographically (to leading order in $1/N$), one considers a dual spacetime whose Euclidean onshell action gives minus
$\log Z(1)$. The replica holographic dual is constructed by considering a boundary theory in which the circle has period $n$ times the length of the original circle. Then $\log Z(n)$ is minus the action $I_E(n)$ for a smooth solution of the bulk field equations in which the circle has a periodicity of $n$ times the original periodicity. Hence we can use holography to rewrite \eqref{lm1} as 
\be
S =  n \partial_n \left [  I_E(n) - n I_E(1) \right ]_{n=1} \label{lm2}
\ee
The second term in \eqref{lm2} is associated with the solution at $n=1$ but with the circle having periodicity of $n$ times the original length; this solution has a conical singularity but the contribution of the conical singularity to the onshell action is not included. 

In \cite{Lewkowycz:2013nqa} the main focus was implicitly asymptotically anti-de Sitter geometries, i.e. solutions of lower-dimensional gauged supergravity theories, in which the bulk actions are Einstein gravity coupled to matter fields.  However, the general arguments given in \cite{Lewkowycz:2013nqa} apply equally to any holographic dual and therefore, in particular, apply to solutions of type IIB supergravity in ten dimensions which asymptote to $AdS_5 \times S^5$. 

In cases where a consistent truncation exists, one obtains the same result for working out the onshell (Einstein frame) action from ten dimensions using the Euclidean continuation of \eqref{iib}
\be
I_E = - \frac{1}{16 \pi G_{10}} \int d^{10}x \sqrt{g} \left [ R  - \frac{1}{2 \cdot 5!} F_{(5)}^2 + \cdots \right ]
\ee
as one does using the five-dimensional Euclidean action:
\be
I_E = - \frac{1}{16 \pi G_5} \int d^5 x \sqrt{g_5} \left [ R(g_5) + \cdots \right ] \label{5daction}
\ee
where the ellipses denote the matter contributions to the consistent truncation. In cases for which no consistent truncation exists, it remains true that the onshell action computed from ten dimensions, by construction, gives the same result as the five-dimensional onshell action. However, when no consistent truncation exist, the ellipses in 
\eqref{5daction}  include the complete tower of Kaluza-Klein modes. 

\bigskip

We now need to argue that \eqref{lm2} localises on a minimal surface and is given by 
\be
S = \frac{\cal A}{4 G_{10}} = \frac{A}{4 G_5}
\ee
where $A$ is the area of the Ryu-Takayanagi surface and ${\cal A}$ is the area of the codimension two minimal surface in ten dimensions. 

Let us first give an argument following the approach of \cite{Lewkowycz:2013nqa}. Let ${\cal M}_n$ be the regular bulk geometry corresponding to $\tau$ being periodic with period $2 \pi n$ and let ${\cal M}_1$ be the geometry with a conical singularity.  Note that the conical singularity extends to the conformal boundary, in contrast to the black hole setup discussed in \cite{Lewkowycz:2013nqa} in which it was localised in the interior of the bulk geometry.

Now the argument given in \cite{Lewkowycz:2013nqa} goes as follows. Consider a smooth offshell configuration with geometry $\tilde{\cal M}_n$ which regularises the conical singularity of ${\cal M}_1$. Away from the fixed point surface the geometry of $\tilde{\cal M}_n$ agrees with that of ${\cal M}_1$ and $\tilde{\cal M}_n$ is chosen such that the offshell configuration differs by order $(n-1)$ from a solution of the equations of motion. Let $\tilde{I}_E(n)$ be the onshell action of the configuration with geometry $\tilde{\cal M}_n$. Since the offshell configuration can always be viewed as a first order variation of an onshell configuration (working perturbatively in the expansion parameter $(n-1)$), its action is equivalent to $I_E(n)$ up to quadratic order in $(n-1)$ and therefore we can replace $I_E(n)$ by $\tilde{I}_E(n)$ in \eqref{lm2}:
\be
S = n \partial_n \left [ \tilde{I}_E(n) - n I_E(1) \right ]_{n=1} \label{lm3}
\ee
Since the geometries only differ at the fixed point set, it is then apparent that this expression localises on the fixed point set. Moreover, the contribution is extensive in the area of the codimension two fixed point set and is proportional to the integral over the cone directions
\be
\int d^2 x \sqrt{g} {\cal R} \sim 4 \pi (1 -n). \label{lm4}
\ee
Thus we can write
\be
S = \frac{\cal A}{16 \pi \mathcal{G}_N} \left ( - n \partial_n \int d^2 x \sqrt{g} {\cal R} \right )_{n=1} = \frac{\cal A}{4 \mathcal{G}_N}, \label{lm5}
\ee
where $\mathcal{G}_N$ is the Newton constant. 

This general argument clearly does not depend on the spacetime asymptotics, and is thus equally applicable to asymptotically $AdS$ and asymptotically $AdS \times S$ geometries. The overall constant of proportionality obtained in \eqref{lm5} can always be fixed by exploiting the CHM map, relating spherical region entanglement entropy to hyperbolic black hole entropy. The latter is given by the standard expression, the area of the horizon (in the Einstein frame metric) divided by $4 \mathcal{G}_N$. 

\bigskip

Starting with \eqref{lm3} we can give a different argument that this expression localises on fixed point sets of the vector $\partial_{\tau}$ using the work of Gibbons and Hawking \cite{Gibbons:1979xm}. We consider the case in which $\partial_{\tau}$ is a Killing vector both on the boundary and in the bulk\footnote{Throughout this paper we have assumed Poincar\'{e} invariance of the dual field theory. In the case of the half space entangling region shown in Figure \ref{fig:replica1} this guarantees that $\partial_{\tau}$ is indeed a Killing vector.}. Since $\tau$ is a circle symmetry we can always write the metric locally as
\be
ds^2 = V (d \tau  + \omega)^2 + V^{-1} ds_B^2 \label{GH0}
\ee
where the scalar $V$ and the one form $\omega$ take values on the base space $B$, which is the space of non-trivial orbits of the circle symmetry. The fibering is trivial if the one form $\omega$ is globally exact; we have implicitly assumed this above. By construction the onshell Euclidean action can be expressed as an integral over $B$:
\be
I_E = \int d \tau d^{D-1} x \sqrt{g} {\cal L} = \beta_{\tau} \int_B d^{D-1} x V^{-1} \sqrt{g_B} {\cal L}  \label{GH1}
\ee
where $\beta_{\tau}$ is the periodicity of $\tau$, $\sqrt{g_B}$ is the base metric determinant and ${\cal L}$ is the onshell Lagrangian. 

The circle symmetry $k = \partial_{\tau}$ has fixed points wherever $V=0$. The action of the symmetry is generated by the antisymmetric matrix $D_{ [M} K_{ N]}$; such matrices have even rank, i.e. rank $(2,4...)$. (The zero rank case would imply that the Killing vector is zero and acts trivially.) When $D_{ [M} K_{ N]}$ has rank $2k$ the action of the symmetry leaves fixed a $(D- 2k)$-dimensional submanifold. Note that when $\omega$ is globally exact the only possible fixed point sets are of dimension $(D-2)$. 

Gibbons and Hawking showed in \cite{Gibbons:1979xm} that for four-dimensional Einstein gravity the onshell action \eqref{GH1} can be expressed as the divergence of a Noether current $J^s$ associated with a dilation symmetry:
\be
I_E = \beta_{\tau} \int_B d^{D-1} x \sqrt{g_B} D_s J^s = \beta_{\tau} \int_{\partial B} d^{D-2} \sigma_s J^s \label{GH2}
\ee
and hence the action localises on the $(D-2)$-dimensional boundary $\partial B$ of the base space. 

The boundary $\partial B$ consists of $(D-2)$-dimensional boundaries surrounding each fixed point together with the spatial boundary at infinity (if $B$ is non-compact). When $\tau$ is the imaginary time, contributions from infinity are associated with conserved charges (mass ${\cal M}$ etc) while contributions from the fixed point sets give the entropy ${\cal S}$:
\be
I_E = \beta_{\tau} {\cal M} + \cdots - {\cal S} \label{GH3}
\ee
where the ellipses denote contributions from additional conserved charges. 

The entropy ${\cal S}$ includes not only the usual area terms but additional contributions associated with a scalar potential $\psi$ dual to the one-form $\omega$, see \cite{Gibbons:1979xm}:
\be
d \psi = V^{2} \ast_3 d \omega \label{GH2z}
\ee 
where the dual is computed on the base space $B$. In four dimensions the fixed point sets are either two-dimensional bolts, characterised by their self-intersection $Y$, or zero-dimensional nuts, characterised by relatively prime integers $(p,q)$. The entropy contributions are then given by 
\be
{\cal S} = \frac{\cal A}{4 G_4} +  \frac{\beta_{\tau}^2}{16 \pi G_4} \sum_{\rm bolts} \psi Y +  \frac{\beta_{\tau}^2}{16 \pi G_4} \sum_{\rm nuts} \frac{\psi}{pq} \label{GH3a}
\ee
where the scalar potential $\psi$ is invariant over a bolt. 

The scalar potential contributions are zero if $\omega$ is globally exact, and then the entropy reduces to the usual form:
\be
{\cal S} = \frac{\cal A}{4 G_4} \label{GH4}
\ee 
with ${\cal A}$ the sum of the areas of $(D-2)$-dimensional fixed point sets. 
The expressions \eqref{GH2} and \eqref{GH3} are believed to apply to Einstein gravity coupled to matter in all dimensions although explicit expressions for the terms in the entropy depending on the one-form $\omega$ are not known in general dimensions. 

In the case at hand, the circle direction is not the imaginary time but the approach of \cite{Gibbons:1979xm} can still be applied, provided that the circle direction is a symmetry. Thus the onshell action can be expressed as 
\be
I_E = \beta_{\tau} \sum_a \Phi_a Q_a - {\cal S}
\ee
where $\Phi_a$ and $Q_a$ are conjugate potential and conserved charge pairs, respectively, and ${\cal S}$ is again associated with fixed point sets of the circle symmetry. By construction we choose $\tilde{\cal M}_n$ to be such that the charge terms cancel between the two terms in \eqref{lm3} leaving
\be
S = [ n \partial_n (n-1)]_{n=1}  {\cal S} = {\cal S},
\ee
i.e. the entanglement entropy is equal to the geometric entropy, which is given by \eqref{GH4} when the fibration in \eqref{GH0} is trivial. 

Note that the derivation using \eqref{GH2} relies on $\partial_{\tau}$ being a Killing vector. While $\tau$ is periodic, it is not necessarily a symmetry direction even for generic entangling regions on flat spatial hypersurfaces of constant time. For example, consider the spherical entangling region $w = R$ in a four-dimensional quantum field theory in the flat background
\be
ds^2 = (dx^0)^2 + dw^2 + w^2 (d \theta^2 + \sin^2 \theta d \phi^2)
\ee
where $x^0$ is the imaginary time. By changing coordinates as
\be
w = R + \tilde{w} \cos \tau \qquad
x^0= \tilde{w} \sin \tau
\ee
to 
\be
ds^2 = d \tilde{w}^2 + \tilde{w}^2 d \tau^2 + ( R + \tilde{w} \cos \tau)^2 ( d \theta^2 + \sin^2 \theta d \phi^2)
\ee
we note that the boundary of the entangling region is at $\tilde{w} = 0$, with $\partial_\tau$ having a dimension two fixed point set at $\tilde{w} = 0$. Here $\tau$ is the circle direction used in the replica trick, but it is not a symmetry. In most previous discussions of holographic entanglement, the circle direction $\tau$ was trivially fibered but not necessarily a symmetry.

On the other hand, the approach of \eqref{GH2} raises the interesting possibility that there may in general be additional leading order contributions to the holographic entanglement entropy, beyond the area of the extremal surface. Suppose that the following metric describes the geometry near the boundary of an entangling region (in a three-dimensional field theory):
\be
ds^2 = d\tilde{w}^2 + \tilde{w}^2 ( d \tau + a(\tilde{w}) d \phi)^2 + b(\tilde{w}) d\phi^2, \label{gh5}
\ee
with $b(0) \neq 0$. Here $\partial_\tau$ is  a Killing vector with a two-sphere fixed point set at $\tilde{w} =0$, which is interpreted as the boundary of the entangling region. Note that for suitable choices of $(a(\tilde{w}),b(\tilde{w}))$ one can obtain \eqref{gh5} as a limit of the Euclidean Kerr-de Sitter metric. 

Given the boundary metric \eqref{gh5} in the vicinity of the entangling region boundary, one can then reconstruct the asymptotic expansion of the 4-dimensional bulk metric:
\be
ds^2 = \frac{d \rho^2}{\rho^2} + \frac{1}{\rho^2} g_{st} dx^s dx^t \label{FG4}
\ee
with 
\be
g_{st} dx^s dx^t =  d\tilde{w}^2 + \tilde{w}^2 ( d \tau + a(\tilde{w}) d \phi)^2 + b(\tilde{w}) d\phi^2 + {\cal O} (\rho^2)
\ee
where terms at order $\rho^2$ can be computed from the curvature of the boundary metric, see  \cite{deHaro:2000xn}. Thus, the fixed point set of $\partial_{\tau}$ is extended to a two-dimensional surface in the bulk. Following the logic above, the associated entanglement entropy should depend not just not on the area of this surface but also on the non-trivial fibration of this circle direction over the surface. From \eqref{FG4} one can deduce that the potential \eqref{GH2z} satisifes
\be
d \psi = \frac{\tilde{w}^3}{\rho^2} \frac{a'}{\sqrt{b}} d \rho + \cdots
\ee
and hence the potential $\psi$ is indeed constant on the surface defined by $\tilde{w}(\rho)$ with $\tilde{w} \rightarrow 0$ as $\rho \rightarrow 0$. Integration of this equation to find the potential and hence apply the formula \eqref{GH3a} would however require the full bulk reconstruction and we postpone this analysis for future work.

\section{Conclusions} \label{eight}

In this paper we have presented evidence that the entanglement entropy computed from top-down \eqref{hrt} is equivalent to that computed using the Ryu-Takanagi formula \eqref{rt}; we showed that the formulae agree in a wide range of examples and used general arguments based on the replica trick. Both formulae, \eqref{rt} and \eqref{hrt}, are applicable to time independent situations. It would be interesting to generalise the analysis of this paper to the covariant holographic entanglement entropy \cite{Hubeny:2007xt} and, in particular, to understand whether contributions associated with non-trivial fibration of the circle coordinate over the entangling region boundary can indeed arise. 

The relationship between the ten-dimensional solution and the lower-dimensional asymptotically AdS solution is in general very complicated. To calculate quantities related to the dual stress energy tensor, one needs to extract the asymptotic form of the lower-dimensional metric, which is related to the ten-dimensional metric by derivative field redefinitions. It is computationally complex to extract the required field redefinitions. The agreement between \eqref{rt} and \eqref{hrt} imposes constraints on the field redefinitions which can be used both to check Kaluza-Klein holography calculations and, in symmetric situations, to infer the lower-dimensional fields, without going through the entire Kaluza-Klein holography procedure. 

In this paper we have focussed primarily on backgrounds which are asymptotic to $AdS_{d+1} \times X$ but the general arguments of section \ref{eight} are equally applicable to any gauge/gravity duality for which the conformal boundary is timelike and the bulk theory is described by Einstein gravity. Thus in particular the top-down entanglement entropy \eqref{hrt} is applicable to top-down realisations of Lifshitz and Schr\"{o}dinger (with one example of the latter being given in section \ref{four}). The formula \eqref{hrt} is also applicable to non-conformal brane dualities \cite{Itzhaki:1998dd}, in the regimes where supergravity is a valid description. 

Our results have implications for the long standing question of how the compact part of the bulk spacetime is reconstructed from field theory data: entanglement entropy tells us about minimal surfaces in the top-down geometry. One could use these surfaces to explore how global features of the top-down geometry are reconstructed.

\section*{Acknowledgments}

We would like to thank Nico Jokela and Kostas Skenderis for useful comments and discussions. 
This work was supported by the Science and Technology Facilities Council (Consolidated Grant ``Exploring the Limits of the Standard Model and Beyond'') and by the Engineering and Physical Sciences Research Council. 
MMT was supported in part by National Science Foundation Grant No. PHYS-1066293 and the hospitality of the Aspen Center for Physics.
We thank the 2015 Simons Center Summer Workshop and the  Galileo Galilei Institute for Theoretical Physics for hospitality and the INFN for partial support during the completion of this work.


\providecommand{\href}[2]{#2}\begingroup\raggedright\endgroup

\end{document}